\documentclass[english]{article}
\usepackage[T1]{fontenc}
\usepackage[latin9]{inputenc}
\usepackage{amsmath}
\usepackage{setspace}
\usepackage{amssymb}
\usepackage{esint}

\makeatletter
\usepackage{babel}
\makeatother

\begin{document}

\title{\textbf{Existence of biological uncertainty principle implies that
we can never find 'THE' measure for biological complexity.}}

\author{\textbf{Anirban Banerji}\\
\textbf{Bioinformatics Centre, University of Pune}\\
\textbf{Pune-411007, Maharashtra, India}\\
\textbf{anirban@bioinfo.ernet.in, anirbanab@gmail.com}\\
}

\maketitle
There are innumerable 'biological complexity measure's. While some
of these measures contradict each other, general patterns emerge from
other attempts to represent biological complexity. Nevertheless, a
single measure to encompass the seemingly countless features of biological
systems, still eludes the students of Biology. It is the pursuit of
this paper to discuss the feasibility of finding one complete and
objective measure for biological complexity. A theoretical construct
(the 'Thread-Mesh model') is proposed here to describe biological
reality. It segments the entire biological space-time in a series
of different biological organizations before modeling the property
space of each of these organizations with computational and topological
constructs. Acknowledging emergence as a key biological property,
it has been proved here that the quest for an objective and all-encompassing
biological complexity measure would necessarily end up in failure.
Since any study of biological complexity is rooted in the knowledge
of biological reality, an expression for possible limit of human knowledge
about ontological biological reality, in the form of an uncertainty
principle, is proposed here. Two theorems are proposed to model the
fundamental limitation, owing to observer dependent nature of description
of biological reality. They explain the reasons behind failures to
construct a single and complete biological complexity measure. This
model finds support in various experimental results and therefore
provides a reliable and general way to study biological complexity
and biological reality.

\section{\underbar{Introduction }:}

Biological complexity measures are many \cite{Edmonds1999}. While
all these measures are useful (because they quantify certain aspects
of the biological systems), in many of the cases, they tend not to
consider the gamut of properties that a complex system is known to
possess in general (emergence, near-neighbor interactions, non-linear
functional dependencies, feedback loops - to name a few) \cite{Hazen2007}.
The words 'complexity measure' or 'complexity index' have become almost
synonymous with some kind of marker for the complex system under observation
\cite{Chiappini2005,Banerji2003}. Although a review of all the methodologies
proposed for the studies of biological complexity is not the objective
behind this work, it assumes importance to capture a glimpse of the
spectrum of significant outlooks prevalent in contemporary studies
on the subject. Such glimpse exposes us to the glaring nature of contradictions
among the existing complexity measures.\\

Description length, in some way or other, forms the basis of most
of the complexity measures \cite{Lofgren1977}. The strict and classical
measure of complexity, namely the KCS definition \cite{Kolmogorov1965,Chaitin1966,Solomonoff1964}
states that for a universal Turing Machine, the KCS complexity of
a string of characters describing it will be given by the length of
the shortest program running on it that generates the description.
However, apart from the theoretical problem of being non-computable
to obtain any reasonable practical estimate\cite{Badii1997}; the
implication of KCS complexity measure, namely, to contain maximum
information the sequence concerned should be absolutely random\cite{Gellmann1994},
contradicts the nature of biological organizations completely. The
case of one\cite{Hinegardner1983} measure of (structural) complexity,
which builds upon the KCS paradigm and counts the number of different
parts that a system contains, highlights the aforesaid contradiction.
The criticisms to this measure are threefold; first, it is difficult
to identify the \char`\"{}parts\char`\"{}; second, the so-called 'C
value paradox'\cite{Thomas1971} and third, ambiguous results originating
from 'coding-non-coding ratio'. To elaborate a little, C value paradox
suggests an absence of correlation between phenotypic complexity with
total size of the genome (even though the plant $Psilotum$ $nudum$
is widely viewed as easy to understand than the plant Arabidopsis,
the former has 3000 times as much DNA; similarly the phenotypic complexity
of a lungfish($\mathrm{C\backsim1.4\times10}^{11}$ base-pairs), has
been counted to be higher than phenotypic complexity of us, the $Homo$
$Sapiens$($\mathrm{C\backsim3.4\times10}^{9}$ base-pairs)). On the
other hand, confounding results arise out of 'coding-non-coding ratio'
too. The use of the number of protein-coding gene as a measure of
biological complexity would make urochordates and insects less complex
than nematodes; and alsos, humans less complex than rice. The inadequacy
of applying this school of thought to compute biological complexity
is discussed by Szathmary \cite{Szathmary2001}.\\

From the paradigm of DNA base-pairs to the realm of ecology, the theme
of construction of a complexity measure for biological systems, only
to be regarded subsequently as either erroneous or ill-suited for
biological considerations, appears recurrent. One approach \cite{Uso2000}
had measured ecological complexity by counting number of synonyms
for a particular process, comparing between models that describe same
set of executed behaviors. However, it was recognized very soon \cite{Barnstad2001}
that not only the information regarding who is interacting with whom
but also the information related to the strength of various (time-dependent)
coupling schemes is what should be taken into account. Soon after,
still another approach bearing a distinct similarity with Hinegardner's
philosophy \cite{Anand2003} surfaced and asked the question \char`\"{}if
diversity is a part of complexity, then should not biodiversity be
a part of biocomplexity?\char`\"{}. Evolution of these ideas clearly
demonstrate that while none of the proposed complexity measures might
be wrong, none of them can encompass all the aspects pertinent to
the system under consideration.\\
\\
Incompleteness of a complexity measure and subsequent renunciation
of it (contradicted by another complexity measure or otherwise) is
nowhere more prominent than in the paradigm of biological sequence
related complexity measures. Even a fleeting glimpse at the linguistic
approaches enables us to appreciate the extent of discord between
them. From the rather simple and particular model of alphabet symbol
frequencies \cite{Wootton1996} to general algorithm concerning clustering
characteristics of cryptically simple sequences \cite{Alba2002},
from popular approaches for evaluation of the alphabet capacity with
the help of combinatorial complexity and linguistic complexity \cite{Milosavljevic1993,Gabrielian1999}
to a complexity measure based on text segmentation by Lempel and Ziv
\cite{Lempel1976} and subsequent modifications \cite{Gusev1999,Chen1999},
from studies in stochastic complexity \cite{Orlov2002} to approaches
related to grammatical complexity \cite{Jimenez2002}; numerous approaches
of linguistic complexity with potential relevance to biological systems
have been explored. With a different onus, identification and characterization
of low text complexity regions that might be functionally important,
was studied by many \cite{Hancock2002,Wan2000,Chuzhanova2000}, where
low-complexity regions have been identified as regions of biased composition
containing simple sequence repeats \cite{Hancock2002,Tautz1986}.
But even in the sphere of this sub-approach, several differences of
opinions prevail. Cox and Mirkin \cite{Cox1997} differs from Tautz
\cite{Tautz1986} in asserting that the stretches of sequences having
imperfect direct and inverted repeats should be also considered as
the sequence with low complexity. A review of many of these methodologies
that had attempted to construct complexity measure for sequence level
biology can be found in a recent work \cite{Abel2005}.\\

Other than the linguistic framework, the information theoretic and
graph theoretic approaches are also used extensively to address the
notion of biological complexity. When the former relied principally
upon the notion of compositional diversity; the later, coupled with
control theoretic tool-set, could investigate structural or topological
complexity of many dynamic systems. To provide some characteristic
examples, at the level of networks, Palsson \cite{Papin2002} defined
a pertinent algebraic structure, the 'extreme pathway', to characterize
its length as the size (complexity marker) of the corresponding flux
distribution map. Considering topological \char`\"{}diversity\char`\"{}
of the assigned graphs as functional flexibility of network concerned,
Hildegard Meyer-Ortmanns \cite{Hildegard2003} had proposed another
useful complexity measure. On a firm graph theoretic note Bonchev
\cite{Bonchev2003} has identified some significant complexity markers
by characterizing the networks with respect to connectedness, subgraph
count, total walk count, vertex accessibility, etc.; before observing
that the information theoretic index can also serve to assess the
compositional complexity of a network. However, innovative as they
are, none of these measures have found consistent usage in the biological
treatise; hinting perhaps at their limitations with respect to general
applicability in biological realm.\\
\\
Apart from these, the other complexity measures with a distinct background
of Physics, have also been proposed. Bennett \cite{Bennett1988} wanted
to circumvent the problems associated with KCS school of complexity
measures by defining a measure based on the degree to which the information
has been organized in a particular object. This method named 'logical
depth' had attempted to measure the time needed to decode the optimal
compression of the observed data. However this was bounded from below
by the magnitude of another complexity measure\cite{Grassberger1986}
which quantifies the minimal information one needs to extract from
the past in order to provide optimal prediction. The complexity measure
due to Wolpert and MacReady \cite{Wolpert1997}, namely 'self dis-similarity',
had attempted to fuse information theory with statistical inference.
It wanted to attribute the variation of spatio-temporal signatures
of systems at different scales (instead of mere cardinality of them)
to the complexity marker for it. But regardless of their theoretical
elegance, none of these aforementioned measures have been considered
by the biologist fraternity for much practical applicability, which
tends to suggest that all these measures have (probably) failed to
distinguish between general biological cases beyond some simple ones.\\
\\
A complexity measure proposed in recent past by Hazen \cite{Hazen2007}
presents itself as one that respects the entanglements of biological
systemic features seriously. This attempt revolves around the measurement
of complexity of a system in terms of the 'functional information';
in other words, measurement of the information required by the system
to encode a specific function. It appears to share similar mathematical
philosophy as another sequence-based complexity measure, namely the
T-complexity \cite{Ebeling2001}. The T-complexity works by counting
the numbers of steps required by an alphabet set to construct a string.
However, it has been proved conclusively \cite{Fei2004} that T-complexity
is rather inefficient to describe the biological complexity, even
when viewed at one-dimensional sequence level merely.\\
\\
The discussion of above helps us to identify two significant problems
with the construction of complexity measures in many cases. First,
the context dependent nature of biological systems is addressed rather
incompletely in many of the complexity measures and second, many of
these measures are constructed on the basis of observer dependent
description of biological reality. Although it is widely agreed upon
\cite{Mahner1997,Kaneko1998,Andrian2006,Marguet2007} that there can't
be a biological complexity measure without consideration of the context
dependence (because function of any biological system itself is context
dependent), many measures do not take it into account \cite{Hinegardner1983}.
The second problem concerns the fact that definition of many of the
complexity measures seem to arise from the observer's choice of, what
he considers, important properties of the system \cite{Fluckiger1995}.
These predilections of observer \cite{Cornacchio1977,Bar-Yam2004}
often revolve around description of systemic property under consideration
from the reference frame of an observer and not from the reference
frame of the system itself. Consequently, the defined complexity measure
tends not to be intrinsic to the system being studied but depends
on extrinsic state (properties) of the observer, as well as on his
preferences to study certain aspects of system concerned.\\
\\
These two extremely important aspects of $context\; dependence$ and
$observer\; dependence$ were however not entirely unaddressed. Gellmann
\cite{Gellmann1996} argued that the description of the ensemble is
also determined by a number of external factors; which depend on who
(the observer) is describing. From a different perspective Adami \cite{Adami2002}
had noted that many of the abstract measures for biological complexity
\char`\"{}do not appear satisfactory from an intuitive point of view\char`\"{}
and proposed a measure ('physical complexity') that owes it's root\textbf{
}to the automata theory but is smart enough to bypass the problems
of KCS measure of complexity by identifying genomic complexity with
the amount of information a sequence stores about it's environment.
The context dependence of this measure is its most interesting feature.
The implicit starting point of Adami's work is to\textbf{ }recognize
that there can be no such thing as biological complexity\textbf{ }in
the absence of context (which incidentally is in contradiction to
the views expressed in the construction of another complexity measure
\cite{Roman1998}), because biological function itself tends to be
context dependent. It has been shown recently that 'physical complexity'
can estimate both structural and functional complexity too, at least
for some particular biological macromolecules \cite{Carothers2004}.
However, as has been noted by Seth \cite{Seth2000}, in Adami's measure
the observer of complexity becomes the environment itself and therefore
enables the measure to assert that it is a measure of the information
about the environment, that is coded in the one dimensional biological
reality, viz. the sequence. The problem with such assertion has been
discussed in details by Seth \cite{Seth2000}. Furthermore, although
being biologically more relevant than many other suggested complexity
measures, 'physical complexity' has been found to be difficult to
evaluate in general practice except for some particular cases. Thus,
although theoretically impressive, this measure too has not found
widespread use amongst the biologist fraternity.\\
\\
\\
Hence it can be clearly seen that even after exhaustive research from
various perspectives, no clear definition of biological complexity
measure has emerged hitherto; as have been admitted in some recent
works \cite{Hazen2007,Hulata2005,Bialek2001}. In fact, as it is evident
from the discussions above, the entire field is burdened by numerous
counterclaims and possible sources of contradictions. Since biological
complexity measures try to represent biological reality, a closer
examination of these contradictory nature complexity measures reveal
the contradictory nature of biological reality, as perceived by observers.
Examples of such contradictions about biological reality are provided
later (Section 2.4.3), with (possible) reasons behind such contradictions.
But before delving into those details, we can note that there are
exists two clear clusters of biological complexity measures in the
confusing ensemble of them. One cluster comprises of complexity measures
that recognize emergence as a property of biological systems and the
other which do not. Since emergence has been conclusively proved to
be an unmistakable feature of any complex adaptive system (true characterization
of biological systems) \cite{Hazen2007,Bar-Yam2004,Ricard2004,Gellmann1994};
we can, from now on, justifiably limit ourselves to only the set of
complexity measures that pay adequate importance to biological emergence.
\\
\\
Acknowledging emergence as an important biological property, a toy
model (the 'Thread-Mesh model') to describe biological reality is
proposed here. With the help of this model it has been proved here
(algorithmically and topologically) that the nature of biological
reality is observer-dependent, time-dependent and context-specific.
Since any biological complexity measure attempts necessarily to depict
the biological reality (in some way or other), the Thread-Mesh model
proves that no biological complexity measure can be constructed that
is objective and complete in its description of biological reality.
Furthermore, it proves that the reason for having so many biological
complexity measures and so many (possible) contradictions in them
is due to the subjective and incomplete views of biological reality,
as captured by the observers.

\section*{2.\underbar{ Description of biological reality with the }}

\section*{\underbar{Thread-Mesh model }:}

Thread-Mesh model (TM model) segments the biological space-time into
a series of different biological organizations, viz. the nucleotides;
amino acids; macromolecules (proteins, sugar polymers, glycoproteins);
biochemical networks; biological cell; tissue; organs; organisms;
society and ecosystem; where these organizational schemes are called
threshold levels. Emergence of a single biological property (compositional
or functional) creates a new biological threshold level in the TM
model. Thus, if any arbitrarily chosen i$^{\text{th}}$ biological
threshold level is denoted as \textbf{TH}$_{i}$, \textbf{TH$_{\text{i+1}}$}
will be containing at least one biological property that \textbf{TH$_{\text{i}}$}
didn't possess. Somewhat similar schemes of identification of biological
threshold levels is neither new \cite{Testa2000,Dhar2007} nor unique
and there can be many other intermediary threshold levels too (considering
the fact that existence of one emergent thread distinguishes \textbf{TH$_{\text{i+1}}$}
from \textbf{TH$_{\text{i}}$}). For example, at high resolution one
can consider the secondary, tertiary, quaternary structures too as
separate threshold levels that exist between amino acids and the proteins.
The basic principles for subsequent discourse, however, are general
and can be applied to any threshold level. Every possible property
that a threshold level is endowed with, is represented by a 'thread'
in the TM model. Thus an environmental property will be called as
an 'environmental thread' in the present parlance. Threads can be
compositional, structural or functional. For example, for the biological
threshold level corresponding to the enzymes (threshold level representing
the macromolecules), one of the compositional threads is the amino
acid sequence; whereas the radius of gyration, the resultant backbone
dipole moment and each of the bond lengths, bond angles, torsion angles
are some examples of structural threads and the values for K$_{\text{m}}$,
V$_{\text{max}}$, K$_{\text{cat}}$ are some examples of it's functional
threads.\\
\\
\textbf{\underbar{2.1) Components of TM model :}}\\
Any biological system to be studied and the observer who is interested
to study some properties of that system, both contribute to the formation
of the thread mesh. Threads in the thread-mesh space of any threshold
level can be classified in 3 types : 

Type 1) The systemic threads, which represent the compositional, structural
and functional properties of any threshold level of only the biological
system under consideration (excluding the pertinent environmental
features that might interact with the treshold under consideration);

Type 2) The environmental threads, which represent all the relevant
properties of environment (structural and/or functional) that can
potentially interact with the systemic threads belonging each and
every threshold level, exhaustively. It is purely biological to expect
that different subsets of the gamut of environmental threads will
be relevant for operations with different threshold levels of the
biological system under consideration (an environmental thread that
is important for helping certain operations at the threshold level
of tissues, might not be relevant at the threshold level of nucleotides
and so on ..)

and 

Type 3) The observer threads, representing the observer (along with
his observational (experimental) tool-set). Although the observer
himself is represented by the threshold level of an individual organism,
the entire set of observer threads can interact with thread set representing
any threshold level of the biological system. It is easy to see that
a partial symmetry results when (in the special case), the observer
observes the 'organismic' threshold level, because\textbf{ }the observer
himself exists at the 'organismic' threshold level (consciousness
of tissue or proteins haven't been reported hitherto). However even
this symmetry will not be complete, because the type-3 threads will
be characterized by threads emanating from experimental tool-sets
too, which the type-1 threads are devoid of.

The myriad possible interactions between first and second type of
threads in thread-mesh space, account for the context specific nature
of biological reality to a great extent. The other kind of context
dependence obviously originates from the nature of interactions between
type 1 threads only.

The relevant question at this point might be : \char`\"{}what is the
general nature of properties that we can measure concentrating on
any one threshold level?\char`\"{} Only the threads representing strong
emergent patterns would be more probable to interact with observer's
thread set and thus will make their presence felt to the observer.
As a result, only some of the properties for a particular threshold
level will be noticed while some other subdued systemic features will
not be known to us.\textbf{\underbar{}}\\
\textbf{\underbar{}}\\
\textbf{\underbar{}}\\
\textbf{\underbar{2.2) Assumptions of Thread-Mesh model :}}\\
\textbf{\underbar{Assumption 1)}}\textbf{ }The systemic properties
will be conserved; that is, no systemic thread can be found that destroys
itself without any trace at the higher threshold levels. \\
In other words, the TM model states that a systemic thread (a systemic
property, structural or functional) representing biological threshold
level \textbf{TH$_{\text{i}}$} should either preserve itself as it
is at the threshold level \textbf{TH$_{\text{i+1}}$}, or it will
merge with some other thread (systemic property or environmental property)
representing \textbf{TH$_{\text{i}}$}, to construct systemic property
representing \textbf{TH$_{\text{i+1}}$}. But no biological property
can vanish from the thread-mesh space. On the other hand, a new systemic
thread can always emerge at any threshold level \textbf{TH$_{\text{i+1}}$}
as an entirely novel one. But once present, the lineage of the property
can always be perceived on the higher threshold levels. \\
\\
\textbf{\underbar{Assumption 2)}}\textbf{ }The systemic thread set
representing any threshold level is constant, but the observer thread
set interacting with it varies with time and context.\\
In other words, the total number of threads that define any threshold
level \textbf{TH$_{\text{x}}$}, say \textbf{N} (all possible compositional
and functional threads of the system along with all possible environmental
threads that have a probability to interact with the systemic threads,
under all possible contexts that the biological threshold might experience);
must be time-invariant for that particular threshold level. This implies
that the volume of the thread-space representing any biological threshold-level
will be constant. Observer's thread set that can interact with \textbf{N},
in contrast, is variable. More interaction of the observer's thread
set with \textbf{N} will account for more completeness of our knowledge
about \textbf{N}, similarly less interaction of the observer's thread
set with \textbf{N} will account for less completeness of our knowledge
about \textbf{N}. \\
This assumption of TM model can thus be stated otherwise as, the intersection
between observer's thread-set and systemic thread-set is not invariant
but is a function of prevailing context and time. In other words,
if the intersection between observer's thread-set and systemic thread-set
under any given context at any instance of time t$_{\text{1}}$ constitutes
a set \textbf{A$_{\text{t1}}$}; and the intersection between observer's
thread-set and systemic thread-set under the same context at any other
instance of time t$_{\text{2}}$ constitutes another set \textbf{A$_{\text{t2}}$};
then in general, $|A_{\text{t1}}|\neq|A_{\text{t2}}|$. However the
assumption asserts further that even if $|A_{\text{t1}}|=|A_{\text{t2}}|$,
it is not probable that \textbf{A$_{\text{t1}}$} and \textbf{A$_{\text{t2}}$}
will be having identical compositions of their thread-set. These different
intersections (context-dependent and time-dependent) between observer's
thread-set and systemic thread-set are the ones that cause the subjective
and contradictory inferences about biological systems. Examples of
how different experimentalists (different observers) can interfere
with biological reality to draw different inferences about it, is
provided in a recent work \cite{Xu2006} at the cellular threshold
level and by another study \cite{Moen2005} at the organismic threshold
level. It is due to this subjective nature of our acquired knowledge
about biological reality that we cannot have a general and complete
measure for biological complexity; but rather will only have to be
content with numerous threshold specific context-dependent complexity
measures. \\
\\
\\
Indeed the recent studies \cite{Dokholyan2001}, \cite{Ding2006}
tend to vindicate the assumption of thread-interactions being context-dependent
and time-dependent. A software, 'Medusa' \cite{Ding2006} attempts
to explore the evolution of a protein fold family in a dynamic manner;
that is, by monitoring the time-dependent changes in sequence and
structure upon random mutations of amino acids. This means, the software
attempts to learn about the nature of functional threads generated
by interactions between three types of threads (first, the thread-set
representing complete amino acid sequence alongside several windows
of varying lengths of that same sequence, second, threads representing
every structural features and third, the compositional threads representing
the random mutations of the amino acids) in a time-variant manner.
The same study finds that a \char`\"{}subtle\char`\"{} change in the
compositional nature of a subset of thread-set representing amino
acid level of biological threshold, result in \char`\"{}distinct packing
of the protein core and, thus, novel compositions of core residues\char`\"{};
depicting how a change in certain set of threads at a particular time
and under appropriate context, can change the entire scheme of interactions
and account for the emergence of one particular property (novel core
residues) at the next level of biological threshold from a pool of
possible properties.\\
\\
\textbf{\underbar{Assumption 3)}} The thread-mesh for any threshold
level is constituted of all the threads representing the systemic
properties, environmental properties and observer (observational mechanisms)
properties and will have a bounded geometry typical of that threshold
level. This assumption points to the definite yet distinctly different
existences of biological organizations. It implies that the pattern
of interactions between biological properties prevalent in any threshold
level must be following a definite pattern that is different (either
subtly or markedly) from the pattern of interactions between biological
properties in another threshold level. However as the threads(properties)
of any threshold level are related\textbf{\underbar{ }}to those of
other threshold levels and the thread-mesh for each of these threshold
levels\textbf{\underbar{ }}have unique\textbf{\underbar{ }}bounded
geometries, interaction between\textbf{\underbar{ }}two threads in
any threshold level can potentially influence the thread interactions
in some other threshold level too.\\
\textbf{\underbar{}}\\
\textbf{\underbar{2.3) Properties of Thread-Mesh model }}\textbf{:}\textbf{\underbar{}}\\
\textbf{\underbar{Property 1)}}\textbf{ :} If the emergent thread
that separates threshold level \textbf{TH$_{\text{i+1}}$} from \textbf{TH$_{\text{i}}$}
is not a case of 'strong emergence' \cite{Crutchfield1994,Ricard2004},
the entire thread set of \textbf{TH$_{\text{i+1}}$} is produced from
\textbf{TH$_{\text{i}}$}(the same can obviously be said about \textbf{TH$_{\text{i}}$}
and \textbf{TH$_{\text{i-1}}$}, and so on). However, although the
compositional lineage exists, the thread set of \textbf{TH$_{\text{i+1}}$}
is independent in its function from function of thread set of \textbf{TH$_{\text{i}}$}.
This functional difference originates due to the existence of different
set of pertinent biological contexts for the threshold levels. For
example, although originating from the genes (\textbf{TH$_{\text{i}}$}),
the proteins (\textbf{TH$_{\text{i+1}}$}) can undergo independent
context driven operations (enzymatic cleavage, aggregation with other
molecules, phosphorylation, glycosylation etc...), which are completely
different from the context-driven operations relevant at the nucleotide
level (for example, due to alternative splicing (a context-specific
operation), almost one third of DNA produces different proteins; -
an operation only pertinent at the nucleotide level). In fact, as
noted recently \cite{Cohen2006}, a protein glyceraldehyde-3-phosphate
dehydrogenase, discovered as an enzyme, has now been identified to
have a function in membrane fusion, micro-tubule bundling, RNA export,
DNA replication and repair, apoptosis, cancer, viral infection and
neural degeneration; which would have been impossible if different
contextual constraints for macromolecular threshold level were not
in place.\\
\\
\textbf{\underbar{Property 2)}}\textbf{ :} Although in certain cases
the absolute number of threads representing threshold level \textbf{TH$_{\text{i+1}}$}
might be less than that of \textbf{TH$_{\text{i}}$} (a process referred
to as 'integration' in a previous study \cite{Ricard2004}); a context-specific
decomposition that respects compositional lineage of thread-set representing
\textbf{TH$_{\text{i+1}}$} will reveal that $(|TH_{\text{i+1}}|-|TH_{\text{i}}|)\geqslant1$.
It is important to note that this inequality expresses an innate fact
about behavior of the systemic threads between biological threshold
level and doesn't involve observer thread set at all.\textbf{\underbar{}}\\
This inequality $(|TH_{\text{i+1}}|-|TH_{\text{i}}|)\geqslant1$ tends
to suggest a deeper biological fact; that is, even in the absence
of an observer, the biological reality can only be talked about in
a threshold-dependent manner. That is, even with the complete knowledge
of the thread-set (compositional, structural, functional) of\textbf{
TH$_{\text{i}}$} under every possible context, the complete set of
biological properties representing \textbf{TH$_{\text{i+1}}$} won't
be\textbf{\underbar{ }}known to us. In fact, the possible limit of
knowledge about biological properties about \textbf{TH$_{\text{i+1}}$}
derived from \textbf{TH$_{\text{i}}$} will always\textbf{\underbar{
}}follow the inequality \begin{equation}
[(f(TH_{\text{i+1}})-f(TH_{\text{i}}))\geqslant1]\end{equation}
Some related works exemplify the impossibility to acquire knowledge
about complete thread-set of \textbf{TH$_{\text{i+1}}$}, from the
knowledge of complete set of threads representing \textbf{TH$_{\text{i}}$}.
These studies clearly point to the fact that to represent the contextual
constraints typical of any \textbf{TH$_{\text{i+1}}$}, novel threads
(compositional and/or structural) are required. The nature of such
novel threads representing the typical contextual (compositional and/or
structural threads) constraints of \textbf{TH$_{\text{i+1}}$} can
not be predicted from the knowledge of thread-set representing \textbf{TH$_{\text{i}}$}.
For example, based upon theoretical calculations that takes into account
the compositional and structural threads in the forms of exact magnitudes
of an individual protein's molecular weight, solute radius and solvent
molar volumes along with appropriate interaction factors (Wilke-Chang
equation and Stokes-Einstein equation), an estimate of diffusion factors
for the proteins was constructed\cite{Iyengar2007}. However the study
finds that such estimates were inadequate when describing the protein's
property space when it is undergoing interactions with other macromolecules
in cytoplasm. The context of cytoplasmic reality demands \char`\"{}appropriate
correction factors\char`\"{} that suitably considers cytoplasmic viscosity,
drag and molecular crowding (all being compositional and structural
threads representing cellular reality). The exact nature of parameter
set representing this \char`\"{}appropriate correction factors\char`\"{}
at any \textbf{TH$_{\text{i+1}}$}, can never be ascertained from
the knowledge of \textbf{TH$_{\text{i}}$}. From a different standpoint,
another study \cite{Hut2000} had also established the impossibility
to create a biological cell on the basis of difficulty to ensure \char`\"{}intrinsic
coherence\char`\"{} (the parameter set representing contextual constraints
at \textbf{TH$_{\text{i+1}}$}) obtained from physical studies between
molecular level entities (\textbf{TH$_{\text{i}}$}). These claims
tend to vindicate the second property of TM model.\textbf{\underbar{}}\\
\textbf{\underbar{}}\\
Findings of $eq^{n}-1$\textbf{\underbar{ }}can be expressed in more
formal and general term with the a theorem, on the nature of acquired
knowledge about any threshold level of any biological system. It can
be stated as :\\
\\
\textbf{\underbar{Theorem-1 : }}\textbf{The very nature of biological
reality makes it impossible for any observer to acquire complete knowledge
about the mutual interaction of biological properties}\underbar{ }\textbf{between
any two adjacent biological threshold levels, representing any part
of biological space-time.}\textbf{\underbar{}}\\
\textbf{\underbar{Proof :}}\\
Let us denote any particular biological property (compositional or
structural or functional) of any arbitrarily chosen biological threshold
level (\textbf{TH$_{\text{i}}$}) by thread $th_{i}$. Similarly let
us denote the lineage of that particular biological property, viz.
$th_{i}$, in the adjacent biological threshold level (\textbf{TH$_{\text{i+1}}$})
by thread $th_{i+1}$. (Examples of such lineages are many. At \textbf{TH}$_{protein}$,
many proteins are found with unfavorable hydrophobic and/or non-polar
residues on their surface. But such unexpected structural feature
only makes sense when one observes that it is those hydrophobic residues
on the surface of the protein that serve as hot-spots for other proteins
to bind \cite{Argos1997}, and subsequently that protein-protein interaction
forms a part of some biochemical pathway at the next biological threshold
level, namely at \textbf{TH}$_{pathway}$).\\
\\
Since every biological property operates within a specified bound
of magnitude (referred to as 'fluctuation' in an earlier study \cite{Testa2000})
we describe the range of magnitude that $th_{i}$ can assume by its
inherent entropy $S\left(th_{i}\right)$, where $S\left(th_{i}\right)=\sum_{i}p_{i}log\frac{1}{p_{i}}$;
($p_{i}=Pr\left(\alpha=\alpha_{i}\right)$, $0\leq p_{i}\leq1$ and
$\sum_{i=1}^{r}p=1$). Similarly we describe the range of magnitude
that the lineage of $th_{i}$, viz. $th_{i+1}$, in (\textbf{TH$_{\text{i+1}}$})
can assume by its inherent entropy $S\left(th_{i+1}\right)$, where
$S\left(th_{j=(i+1)}\right)=\sum_{j}q_{j}log\frac{1}{q_{j}}$; ($q_{j}=Pr\left(\beta=\beta_{j}\right)$,
$0\leq q_{j}\leq1$ and $\sum_{j=1}^{s}q=1$).\\
To describe the extent of effect $th_{i}$ has on $th_{i+1}$, we
resort to conditional probability. Denoting the entire expected extent
of the effect of $th_{i+1}$ due to entire $th_{i}$ as $\Gamma_{j}$,
a part of the same as $\beta_{j}$ $\left(\sum_{j}\beta_{j}th_{j}=\Gamma_{j}\right)$,
and the part of $th_{i}$ that accounts for $\beta_{j}$ as $\alpha_{i}$
($\left(\sum_{i}\alpha{}_{i}th_{i}=\sigma_{i}\right)$, where $\sigma_{i}$
accounts for $\Gamma_{j}$); we have :\\
\begin{equation}
S\left(th_{i}|\beta_{j}\right)=\sum_{i}Pr\left(\alpha_{i}|\beta_{j}\right)log\frac{1}{Pr\left(\alpha_{i}|\beta_{j}\right)}=\sum_{i}Q_{ij}log\frac{1}{Q_{ij}}\end{equation}
\\
The parameter $Q_{ij}$ represents the state of (\textbf{TH$_{\text{i+1}}$})
when it is aware that $\beta_{j}$ extent of the property $th_{i+1}$
is operative, but does not possess the entire information content
about the qualitative and quantitative nature about $\alpha_{i}$
extent of $th_{i}$, that is causing $th_{i+1}$ to behave in the
way it is doing. Similarly, by $P_{ij}$, we can describe state of
(\textbf{TH$_{\text{i}}$}) when (\textbf{TH$_{\text{i}}$}) is aware
that $\alpha_{i}$ extent of $th_{i}$ is operative in causing $\beta_{j}$
extent of effect on $th_{i+1}$, but does not possess the entire information
content about the qualitative and quantitative nature of $th_{i+1}$.
Eq$^{n}$-2, by itself, describes the uncertainty in (\textbf{TH$_{\text{i+1}}$})
about the qualitative and quantitative nature of $th_{i}$. The joint
probabilities, say $J_{ij}$, describing the state of an observer
(obviously not a part of the system), attempting to know the qualitative
and quantitative extent of both $\alpha_{i}$ and $\beta_{j}$ can
be described by simple Bayesian structure as :\\
$\forall\: i,j\::\quad p_{i}P_{ij}=Pr\left(\alpha_{i}\right)Pr\left(\beta_{j}|\alpha_{i}\right)=Pr\left(\alpha_{i},\beta_{j}\right)=Pr\left(\beta_{j}\right)Pr\left(\alpha_{i}|\beta_{j}\right)=q_{j}Q_{ij}=J_{ij}$,\\
hence \begin{equation}
Q_{ij}=\frac{p_{i}}{q_{j}}P_{ij}\end{equation}
\\
On averaging over all the $\beta_{j}$s and using $q_{j}Q_{ij}=J_{ij}$,
we derive the expression for equivocation of $th_{i}$ w.r.t $th_{j=(i+1)}$
:\\
\begin{equation}
S\left(th_{i}|th_{j=(i+1)}\right)=\sum_{j}q_{j}S\left(th_{i}|\beta_{j}\right)=\sum_{j}q_{j}\left(\sum_{i}Q_{ij}log\frac{1}{Q_{ij}}\right)=\sum_{i}\sum_{j}J_{ij}log\frac{1}{Q_{ij}}\end{equation}
\\
Eq$^{n}$-4 describes the average uncertainty in (\textbf{TH$_{\text{i+1}}$})
about $th_{i}$ when $th_{i+1}$ is operative.\\
\\
The set of argument described by eq$^{n}$-2 and eq$^{n}$-4 w.r.t
(\textbf{TH$_{\text{i+1}}$}) holds true as mirror image w.r.t (\textbf{TH$_{\text{i}}$})
and can be described as eq-5 and eq-6, respectively as:\\
\begin{equation}
S\left(th_{j=(i+1)}|\alpha_{i}\right)=\sum_{j}Pr\left(\beta_{j}|\alpha_{i}\right)log\frac{1}{Pr\left(\beta_{j}|\alpha_{i}\right)}=\sum_{j}P_{ij}log\frac{1}{P_{ij}}\end{equation}
\\
\\
and averaging over all the $\alpha_{i}$s and using $p_{i}P_{ij}=J_{ij}$,
we derive the expression for equivocation of $th_{j=(i+1)}$ w.r.t
$th_{i}$ as :\\
\begin{equation}
S\left(th_{j=(i+1)}|th_{i}\right)=\sum_{i}p_{i}S\left(th_{j}|\alpha_{i}\right)=\sum_{i}p_{i}\left(\sum_{j}P_{ij}log\frac{1}{P_{ij}}\right)=\sum_{i}\sum_{j}J_{ij}log\frac{1}{P_{ij}}\end{equation}
\\
\\
An observation process that attempts to retrieve qualitative and quantitative
information regarding the lineage of the properties under consideration,
will henceforth be subjected to an average uncertainty given by the
joint entropy :\\
\begin{equation}
S\left(th_{i},th_{j}\right)=\sum_{i}\sum_{j}Pr\left(\alpha_{i},\beta_{j}\right)log\frac{1}{Pr\left(\alpha_{i},\beta_{j}\right)}=\sum_{i}\sum_{j}J_{ij}log\frac{1}{J_{ij}}\end{equation}
\\
Since the mode of information exchange between any two arbitrarily
chosen biological threshold levels $th_{i}$ and $th_{j}$ (and vice-versa),
is farthest from being inependent, using $p_{i}P_{ij}=J_{ij}$, we
arrive at :\\
\begin{equation}
S\left(th_{i},th_{j}\right)=\sum_{i}\sum_{j}J_{ij}log\frac{1}{p_{i}}+\sum_{i}\sum_{j}J_{ij}log\frac{1}{P_{ij}}\end{equation}
\\
$\because$ $\sum_{j}J_{ij}=p_{i}\:\forall i$, hence \\
\begin{equation}
S\left(th_{i},th_{j}\right)=\sum_{i}p_{i}log\frac{1}{p_{i}}+\sum_{i}\sum_{j}J_{ij}log\frac{1}{P_{ij}}=S\left(th_{i}\right)+S\left(th_{j}|th_{i}\right)\end{equation}
\\
Neither the information regarding the magnitude of the parameter $S\left(th_{j}|th_{i}\right)$
nor the same about the qualitative nature of it can be retrieved by
studying the entire thread-set of either (\textbf{TH$_{\text{i}}$})
or (\textbf{TH$_{\text{i+1}}$}), exhaustively.\\
Hence the proof.\textbf{\hfill{}Q.E.D}\\
\\
\\
\textbf{\underbar{}}\\
\textbf{\underbar{Property 3)}}\textbf{ :} The mapping between threads
representing \textbf{i}$^{\text{th}}$ threshold level \textbf{TH$_{\text{i}}$}
with that of another threshold level \textbf{TH$_{\text{i+r}}$} follows
a many-to-many mapping scheme. For example, the mapping from genotype
(DNA) to phenotype (organism) is marked with\textbf{ }significant
redundancy from either side. Different genotypes can map to the same
phenotype; for example, different codons (DNA nucleotide triplets,
representing threads of the nucleotide threshold level) can code for
the same amino acid; ensuring that the genotype can change (a nucleotide
can mutate) without changing the phenotype. On the other hand, the
same genotype can result in different phenotypes, due to different
environmental conditions during development (different time-variant
and context-dependent interactions between systemic and environmental
threads). It should be mentioned here that a case of one-to-many mapping
scheme can also take place in certain situations (interaction-dependent
differentiation rules for stem-cell\cite{Furusawa1998}); however
that will only be a special case of many-to-many mapping.\textbf{\underbar{}}\\
\textbf{\underbar{}}\\
\textbf{\underbar{}}\\
\textbf{\underbar{Property 4)}}\textbf{ :} For any threshold level,
only a subset of possible thread interactions (between systemic and
environmental threads) can give rise to emergent threads at the higher
threshold levels. In other words, the geometry of thread association
defines the production of an emergent thread. For example, it has
been found that from a superset of possible interactions amongst threads
representing biological threshold of amino acids, only a certain subset
can give rise to stable secondary structures and tertiary symmetries
\cite{Li1996}. This is a clear act of emergence, where the emergent
threads are that of electrostatic and thermodynamic stability of the
structures. The relevance of this property can be vindicated by its
top-down counterpart drawn from a study \cite{Laughlin2000} of physical
laws; although the bottom-up approach involving biological causality
was not mentioned there.\\
\\
\\
\textbf{\underbar{2.4) Efficient descriptions with TM Model}}\textbf{}\\
\textbf{\underbar{2.4.1) Description of context-dependence and time-dependence }}

\textbf{\underbar{in biological systems :}}\textbf{ }\\
Time dependence and context dependence do not mean the same; while
time-dependence attempts to capture the observed effect due to same
biological context at different instances of time, context-dependence
attempts to represent the relevant biological context used for the
description of biological process under consideration. The level of
antibody production and cell proliferation of animals treated with
6-hydroxydopamine serves as an interesting example \cite{Kohm1999}
of how the acquired knowledge of biological reality can change as
a result of time-dependent and context-dependent interactions between
systemic threads and observer threads. In this case \cite{Kohm1999},
involvement of various biological threshold levels under observation
had resulted in producing very many types of contexts. In fact, the
patterns seen from the observed facts reveal numerous schemes of different
intersections between interacting thread sets. While some studies
\cite{Besedovsky1979,Williams1981,Kruszewska1995}, have reported
enhancement in antibody production and cell proliferation, another
group of findings tends to reveal a suppression \cite{Hall1982,Livnat1985,Madden1989}
of the same; while still another experiment \cite{Miles1981} reports
\char`\"{}no change\char`\"{} in the level of antibody production.
Assuming none of the results were wrong, reasons behind obtaining
such contradictory results can be understood by TM model. Although
the biological system (and possibly the environment) was kept invariant,
the differing results have originated due to observer's reading of
subtle changes in the composite biological contexts. Each of these
different contexts had provided the favorable conditions of it's own
for different subsets of threads to interact. In order to study the
effect on the threads of any threshold level \textbf{TH$_{\text{i}}$}
involved in the process in a minimal two-context scenario, we consider
context$_{\text{i1}}$ as one which provides suitable condition for
thread-set \textbf{TH$_{\text{i1x}}$} to interact; and context$_{\text{i2}}$,
as one which provides suitable condition for thread-set \textbf{TH$_{\text{i2x}}$}
to interact. Both \textbf{TH$_{\text{i1x}}$} and \textbf{TH$_{\text{i2x}}$}
represent the systemic thread-sets and are independent of observer.
Since neither the cardinality of \textbf{TH$_{\text{i1x}}$} and \textbf{TH$_{\text{i2x}}$},
nor the composition of them are guaranteed to match, different scheme
of interactions, even in the absence of the observer, will account
for emergence of different threads in \textbf{TH$_{\text{i+1}}$}.
The observer thread-set, whenever interacting with either of \textbf{TH$_{\text{i1x}}$}
and \textbf{TH$_{\text{i2x}}$} will produce different intersections,
and in general, these intersections will neither have their cardinality
matched, nor will there be a consensus in their composition. Since
it is not probable to have a match between two-step intersections
(first, between \textbf{TH$_{\text{i1x}}$} and \textbf{TH$_{\text{i2x}}$}
and second, between either of them and the observer), we will always
acquire (potentially) contradictory knowledge about biological systems.
Thus, even in the case where all the observations (experiments) are
correct, the disparity in observed patterns (experimental results)
will be present, as have been proved in an exhaustive experimental
work \cite{Wahlsten2003}. Due to this subtle (or overt) interaction
with the observer thread-set, the systemic thread-set will only be
perceived by different intersections between the two and therefore
the objective biological reality will not be known.\\
\\
\textbf{\underbar{2.4.2) Description of biological emergence :}}\textbf{
}\\
Due to its architectural characteristics, the TM model explains emergence
naturally. It can easily differentiate between two broad kinds of
emergences as mentioned in some previous studies \cite{Bedau1997,Bar-Yam2004}.
The weak emergence is exemplified by threads (biological properties)
that represent \textbf{TH$_{\text{i+1}}$}, who arise from the interactions
of the threads present at \textbf{TH$_{\text{i}}$}, \textbf{TH$_{\text{i-1}}$},
\textbf{TH$_{\text{i-2}}$}, and so on,\textbf{ }\cite{Odell2002}.
There can as well be strong emergence, exemplified by novel threads
at \textbf{TH$_{\text{i+1}}$} that are neither predictable nor deducible
from threads representing preceding threshold levels \textbf{TH$_{\text{i}}$},
\textbf{TH$_{\text{i-1}}$}, \textbf{TH$_{\text{i-2}}$} \cite{Crutchfield1994,Ricard2004}.
The TM model states that there should be at least one emergent thread
(emergent property, be it an example of week emergence or strong emergence)
to distinguish any \textbf{TH$_{\text{i+1}}$} from \textbf{TH$_{\text{i}}$}.
For example, protein fold family formation \cite{Ding2006} can be
recognized as a case of weak emergence (because time-dependent thread
interactions between structural threads at protein(\textbf{TH$_{\text{i}}$})
level and compositional threads representing (randomly mutated) amino
acids(\textbf{TH$_{\text{i-1}}$}) level are responsible for emergence
of a certain protein fold at the threshold level representing various
protein-folds(\textbf{TH$_{\text{i+1}}$})). Due to extreme context-dependence
and nonlinearity in the thread interactions, at times a cause initiated
at any \textbf{TH$_{\text{i-r}}$} can exhibit pronounced weak-emergence
at any higher order threshold level. For example, the triggering of
the eclosion hormone (interaction of some functional thread at \textbf{TH$_{\text{proteins}}$})
had been reported to initiate a sequence of events, which ultimately
results in the emergence of the moth (\textbf{TH$_{\text{organism}}$})
from the pupal exuviae \cite{Truman1973}. On the other hand, the
formation of functional biochemical network in several branches of
molecular cell physiology can be identified to exemplify the strong
emergence \cite{Boogerd2005}. \\
\\
However, it might not always be easy to identify the emergent threads,
because of the low resolution viewing of the threshold level. For
example, it was known that during cell cycle (\textbf{TH$_{\text{Cell}}$})
the G1 to S transition, under any given growth condition, is characterized
by a requirement of a specific and critical cell size, PS. However,
it has been found recently \cite{Barberis2007} that the creation
of PS is itself an emergent property, where the constraints on cell
size only serve as the lower bound of cardinality of compositional
thread set necessary to create the emergent phase PS.\\
\\
Since biological systems are open systems, many (but not all) biological
threshold levels interact with environment. In fact, environmental
threads that influence systemic threads for any threshold level, form
a part of biological context. An example of how environmental threads
at \textbf{TH$_{\text{Nucleotide}}$}can cause emergence on the \textbf{TH$_{\text{Cell}}$}
can be found in recent literature \cite{Barcellos2008}; where it
has been shown that cancer in an organism can be considered as an
emergent phenomenon of genotype (viz., DNA) perturbed through radiation
exposure. Similar observation has been made elsewhere too \cite{Glade2004};
where it has been found that \char`\"{}under appropriate in vitro
conditions\char`\"{} (which implies, only in the presence of certain
environmental threads) microtubules (represented by thread set belonging
to the threshold level sub-cellular macromolecular complex) form dissipative
structures that not only shows self-organization but also displays
emergent phenomena.\\
\\
Although it is difficult to estimate the number of interacting threads
to achieve emergence, in some of the cases, it has been calculated.
For example, based on Landauer's principle it has been calculated\cite{Landauer1967,Davies2004}
that to ensure the onset of emergence of functional proteins (each
aspect of protein functionality denotes one functional thread at \textbf{TH$_{\text{i+1}}$})
from an ensemble of amino acids (each amino acid constitutes a structural
thread at \textbf{TH$_{\text{i}}$}), the inequality 60 < n < 92 should
hold, where n denotes the number of amino acids. A similar calculation
argues that there should be at least 200 base-pairs (i.e.; 200 structural
threads at the nucleotide threshold level) to ensure the emergent
features of a functional gene \cite{Davies2004}.\\
\\
\textbf{\underbar{2.4.3) Description of observer-dependent nature
of biological reality :}}\textbf{ }\\
There are myriad examples of contradictions about acquired knowledge
of biological systems, studied under the same biological contexts.
Here we will enlist only some of them to highlight the incomplete,
subjective and observer-dependent nature of acquired knowledge about
biological reality.

\begin{description}
\item [{Ex.1)\textmd{In}}] a study \cite{Levchenko1997} involving Clp
family of chaperones, the presence of PDZ-like domains in the carboxy-terminal
region of ClpX was reported. However, another study \cite{Neuwald1999}
within a few days, had not only reported the absence of that entire
PDZ-like domain, but also failed to find any significant similarity
to multiple alignment profile of PDZ domains. The cause for this contradiction
can be attributed to different (almost mutually exclusive) compositions
of sets of intersections between observer's thread set with the systemic
thread set.
\item [{Ex.2)\textmd{The}}] msx homeodomain protein is a known downstream
transcription factor of the bone morphogenetic protein (BMP-4) signal
besides being an important regulator for neural tissue differentiation.
A recent study \cite{Ishimura2000} has reported that both BMP-4 and
Xmsx-1 have failed to inhibit the neurulation of ectodermal tissue
that was combined with prospective dorsal mesoderm;\textbf{ }although
the\textbf{ }amount of injected RNA was sufficient for inhibiting
neurulation in the single cap assay. This result is in direct contradiction
with the results obtained from some other studies\textbf{ }about the
same biological process involving the same biological entities and
the same systemic contexts. These studies have reported that the BMP-4
signal is sufficient for the determination of neural cell fate \cite{Sasai1995,Suzuki1995,Wilson1995,Xu1995}.
The cause behind this subjectivity, therefore, can solely be attributed
to different compositions for sets of intersections between observer's
thread set with the systemic thread set.
\item [{Ex.3)\textmd{In}}] the field of determination of phylogeny of Rhizobium
galegae by genome sequencing (nucleotide threshold level), an interesting
case can be observed. Upon sequencing 260 bp of the 16S gene, two
independent reports \cite{Young1991,Nour1994} had inferred that Rhizobium
galegae is closely related to mesorhizobia. However, upon examining
800 bp (instead of 260 bp), another study \cite{Terefework1998} had
inferred that the previous assertion regarding phylogenetic profiling
of Rhizobium galegae, was wrong. Such subjective knowledge about biological
reality is acquired because the possible scopes of interaction of
observer's thread set with the systemic threads were different. In
one case the observer's thread set could interact with 800 bp (representing
a part of systemic thread set of compositional nature) whereas in
the other case, the possibility of interactions were limited by a
small number of systemic thread set (compositional), viz. 260 bp.
The qualitative changes in biological context due to the presence
of 540 bp (800 bp - 260 bp) could not obviously be taken care of by
two previous studies \cite{Young1991,Nour1994}. As a result of a
squeezed systemic space with merely 260 compositional threads and
the corresponding functional threads, possibility of interaction with
observer's thread set was reduced substantially, leading to a contradiction.
\end{description}
While this example demonstrates clearly that constriction of the systemic
thread space under observation (a process represented by the observer
thread set) reduces the probability of the interaction and can eventually
lead to wrong inferences about biological reality, it is clear that
the same logic holds for an expanded systemic thread space under observation
too. In fact, to observe any biological process of interest, experiments
are carefully designed to ensure the presence of certain systemic
threads and not others. The experimental design ensures a suitable
size of the systemic thread space; which, when interacting with the
observer's thread space, reveals certain facets of a biological process.
Although a squeezed or expanded systemic thread set might not change
the underlying biological process under observation (interaction of
systemic threads with the observer's thread set under a given context),
it can account for less or more bulk of information than the observer
can handle. Thus, the process of observation can always be considered
subjective and incomplete; because the set representing intersection
of systemic thread set and observer thread set - will always be different
(with respect to cardinality of the set, as well as with respect to
composition of the set), even under the same context.\\
\\
\\
\textbf{\underbar{2.5) Mathematical framework of TM model }}\textbf{:}\\
Despite the fact that there have been attempts to develop mathematical
models to describe emergence \cite{Rasmussen1995,Bonabeau1997,Bar-Yam2004,Ricard2004};
an universally accepted mathematical model to describe biological
emergence has still not been found \cite{Cohen2006}. Here with the
TM model we propose a mathematical template that attempts to suitably
describe biological emergence. More importantly, it is proved here
that the nature of biological reality is observer-dependent (hence
subjective). Owing to this inherent observer-dependent knowledge of
biological reality, contradictions arise in the descriptions of it
and it is due to the observer predilection that there are so many
complexity measures to describe the biological reality. Further, it
is due to this innate nature of biological reality that it will not
be possible to construct a biological complexity measure that is objective.\\
\textbf{\underbar{Section : 2.5.1)}}\\
Let us define :\\
\textbf{(Def$^{\text{n}}:$1) T} : The superset of all the threads
(compositional, structural and functional properties) that completely
represent each of systemic features, environmental features and observer
(features of observational mechanism that interacts with the biological
system) at any threshold level \textbf{TH$_{\text{i}}$}.\\
\textbf{(Def$^{\text{n}}:$2) }$\mathbf{\tau}:\left\{ \tau\right\} _{i=1}$where
$\tau\in T$; are the threads representing all the systemic-environmental
(systemic and environmental) properties at \textbf{TH$_{\text{i}}$}.\\
\textbf{(Def$^{\text{n}}:$3) }$o:\left\{ o\right\} _{i=1}$where
$o\in T$; are the threads representing properties of the observer
(and observational mechanisms) at \textbf{TH$_{\text{i}}$}, that
can possibly interact with \textbf{$\tau$}.\\
Thus the interaction between observer and systemic-environmental thread-set
can be described in the broadest terms with the entire spectrum of
threads that represent them as :\\
$\Omega_{1}(\mathbf{\tau})\cup\Omega_{2}(\mathbf{o})$ , where both
$\Omega_{1}$ and $\Omega_{2}$ are context-dependent and time-dependent
functions that model the favorable conditions for the observer threads
to interact with the systemic-environmental threads.\textbf{}\\
\textbf{(Def$^{\text{n}}:$4) MI} : An index set of mutually interacting
pair of threads (mutually interacting biological properties). The
elements of this set; viz. the pair of threads, are not pairwise disjoint
at any instance of time.\\
Hence formally, if $|MI|=\xi$ and for each $i\in MI$ let $\lambda_{i}$
be a minimally interacting set, then :\\
$\left\{ \lambda_{i}:\forall i\in MI\right\} $, $\lambda_{i}=^{\xi}C_{2}$
and $\forall\lambda_{i},|\lambda_{i}|=2$\\
Since the thread-mesh for any threshold level is bounded in its geometry
(by assumption), any change of thread coordinate will definitely influence
coordinate of other threads (either subtly or markedly). Biologically
this implies that every biological property (compositional, structural,
functional) has some influence (however subtle or pronounced) on all
the other biological properties of the same threshold level and potentially
on some properties belonging to other threshold levels too.\\
Thus the threads can not be pairwise disjoint and we have : $\underset{i\in MI}{\cap}\lambda_{i}\neq\phi$\\
For a general case, we'll have : $\lambda_{i}=^{\xi}C_{\eta},\eta>2$
and $\eta\longrightarrow\xi$;\\
where $\eta=\eta$(time, biological context) and clearly $|\lambda_{i}|>2$.\textbf{}\\
\textbf{(Def$^{\text{n}}:$5) $\mathbf{\tau^{+}}$ }and \textbf{$\mathbf{\tau^{-}}$}:
$\mathbf{\tau^{+}}\subset MI$, represents the set of (context-dependent)
interacting thread-pairs that help the formation of an emergent thread
(this emergence, obviously, is an example of weak emergence; strong
emergence does not result from a causal lineage \cite{Crutchfield1994,Bar-Yam2004}).
The \textbf{$\tau^{-}$}, where\textbf{ }$\mathbf{\tau^{-}}\subset MI$,
represents the set of (context-dependent) interacting thread-pairs
that do not contribute to the emergence of that particular thread.
Since $\mathbf{\tau^{+}}$ and \textbf{$\mathbf{\tau^{-}}$} are context-dependent
and time-dependent, $\mathbf{\tau^{+}}\cap\mathbf{\tau^{-}}\neq\mathbf{\phi}$;
which makes perfect biological sense. For example, roughness of the
patches of protein surface might not have a great dependence on resultant
dipole moment of the protein arising out of its main-chain and vice
versa. Therefore the threads representing these two properties might
not always interact. However under certain conditions, say in a highly
polar environment with low dielectric constant, the dipole moment
of the protein backbone might influence the surface geometry and henceforth
the finer aspects of surface topology too, and in such a case the
threads of the aforementioned properties will interact with each other.
Thus while \textbf{$\mathbf{\tau^{+}}$ }and \textbf{$\mathbf{\tau^{-}}$}
can not be regarded as mutually exclusive, their (possible) intersection
depends on particular biological contexts.\\
If we define $\mathbf{|}\mathbf{\tau^{+}|_{time,context}}=\omega$,
the quantitative change in the interacting thread population that
cause weak emergence can then be represented as: $\frac{d}{dt}(\mathbf{\tau^{+}})=\omega f(\omega)$,
where any suitably found $f(\omega)$ will represent the functional
characteristics of time-dependent, context-specific population of
interacting threads, such that $\forall\omega_{1},\omega_{2};\omega_{1}<\omega_{2}\Rightarrow f(\omega_{1})<f(\omega_{2})$.\\
This is biologically appropriate structure. For example, it has been
found in the case of spatio-temporal cytoplasmic organization that
an increase or decrease of the glycolytic flux is induced by an increase
or decrease of polymeric microtubular proteins, as instances of emergence
within a metabolic network \cite{Aon2004}.\\
If $|\frac{d}{dt}(\mathbf{\tau^{+}})|>0$ , we denote $\frac{d}{dt}(\mathbf{\tau^{+}})$
as \textbf{H}. However, even if \textbf{H} exists, the necessary condition
for weak emergence for the interacting threads representing \textbf{TH$_{\text{i}}$}
is satisfied only when \textbf{TH$_{\text{i+1}}$} holds at least
1 thread (a property, be it compositional, structural or functional)
present in \textbf{H.} In that case the inequality of (eq$^{\text{n}}$-1)
will be satisfied. This inequality (along with eq$^{n}$-9) suggests
a possible limit of our knowledge of biological properties representing
any biological threshold level\textbf{ TH$_{\text{i+1}}$}, even if
we (ideally) know the complete set of biological properties that characterize
\textbf{TH$_{\text{i}}$}. \\
\\
We can now proceed to prove that nature of biological reality is observer-dependent.
However before than that it is important to derive an idea of the
topological nature of biological space with the help of TM model.\\
\\
\textbf{\underbar{Section : 2.5.2)}}\textbf{\emph{\underbar{}}}\\
\textbf{\emph{\underbar{2.5.2.1) Topological nature of biological
space }}}\textbf{\emph{:}}\textbf{\emph{\underbar{}}}\\
Biological space can be described in terms of the generalized coordinates
of 'systemic' properties (compositional, structural and functional)
of the biological system under consideration along with the relevant
set of property of the environment, that exerts definite influence
on the systemic properties. Each one of these properties (systemic
and environmental) can be represented by structures called 'threads'.
Hence an ensemble of interactive biological properties will give rise
to a mesh of threads, the 'thread-mesh'. Such description of biological
space is abstract but is advantageous in its being independent of
any particular coordinate system. Furthermore, it describes completely
what a system is comprised of (compositional and structural threads)
and what the system is capable of performing (functional threads)
and how (interactions amongst the pertinent threads). \\
The topological properties of thread-mesh space for any biological
threshold level offer interesting insights. The two components of
such space can be identified as, first, the 'biological-system thread-mesh'
(exhaustive set of compositional, structural and functional properties,
describing any arbitrarily chosen \textbf{i$^{\text{th}}$} biological
threshold level, say \textbf{Th$_{\text{i}}$}; including the relevant
environmental properties) and second, the 'observer thread-mesh' (exhaustive
set of compositional, structural and functional properties of the
observer, which includes the subset of properties required to study
\textbf{Th$_{\text{i}}$}). We assume that each of these biological
properties can be represented by some suitable mathematical functions.\\
\\
Resorting to functional analysis, if we tend to represent these functions
by vectors; we arrive at an abstract vector space to represent biological
properties at any biological threshold level. Such representation
of biological property space (thread-mesh) enables us to define a
\char`\"{}distance\char`\"{} between two functions (any two biological
properties) by $d(p_{\text{1}},p_{\text{2}})=||p_{\text{1}}-p_{\text{2}}||$.
This distance will correspond to the difference between the nature
of biological properties, when these properties are represented by
threads. Two closely related biological properties will have a small
distance between them in the abstract thread-mesh coordinate space.
For example, for \textbf{TH$_{\text{protein}}$}, the distance between
the thread representing interior dielectric constant and the thread
representing probability of interior salt-bridge formation will always
be less than the distance between the thread representing interior
dielectric constant of proteins and the thread representing the shape
of the proteins. The distance between two threads can always be measured
and since any biological property is dependent upon other biological
properties, the distance $d(p_{\text{1}},p_{\text{2}})$ between any
two biological properties of the same threshold level $\left(d(p_{\text{1}},p_{\text{2}})=||p_{\text{1}}-p_{\text{2}}||\right)$
can be considered complete. This turns the thread space (biological
property space) representing \textbf{TH$_{\text{i}}$}, into a metric
space. Also since the threads, who are functions that represent biological
properties, can be associated with their respective lengths (norms);
the abstract vector space of thread-mesh can be considered as a normed
vector space. The effect of any external influence on a vector (biological
property) of this normed vector space can be represented by addition
and multiplication of a scalar variable to that vector. A scaler addition
or multiplication scheme increases or decreases the weightage of a
thread in any thread-interaction. This is necessary, because all the
biological properties assume weightages with respect to the changing
contexts and can't be considered as having invariant importance under
all circumstances.\\
Thus we notice that : \\
\textbf{1)} any biological property representing any biological threshold
level is dependent on all the other properties representing the same
threshold level. Which implies inner-products can be defined on all
the threads (biological properties),\\
\textbf{2)} context-dependent importance of any property (a biological
property modeled as a vector in the normed abstract vector space)
can be modeled by addition and multiplication by a scalar, \\
and\\
\textbf{3)} Cauchy convergence exists amongst the properties representing
any threshold level (for example, the electrostatic interactions (\textbf{I})
between macromolecules in cytoplasm can be broadly modeled as interactions
between set of relevant threads (\textbf{V}), given by : \textbf{th$_{\text{1}}$},
a thread that represents interactions between charges; \textbf{th$_{\text{2}}$},
thread that represents interactions between dipoles(for molecules
without inversion center); \textbf{th$_{\text{3}}$}, representing
interactions between quadrupoles(for molecules with symmetry lower
than cubic); \textbf{th$_{\text{4}}$}, representing interactions
between permanent multipoles; \textbf{th$_{\text{5}}$}, thread representing
induction(between a permanent multipole on one molecule with an induced
multipole on another); \textbf{th$_{\text{6}}$}, that represents
London dispersion forces and \textbf{th$_{\text{7}}$}, a thread that
represents electrostatic repulsions(to prevent macromolecular collapse).
If we consider a local base \textbf{L} for the thread-set of electrostatic
interactions \textbf{I} about any suitably chosen central point (say,
0); then for sequence of threads \textbf{th$_{\text{s}}$} for all
the threads \textbf{V} of \textbf{L}, for some number $\varepsilon$,
whenever any $n,m>\varepsilon$; it will be ensured that \textbf{th$_{\text{n}}$}
- \textbf{th$_{m}$} is an element of \textbf{V}. Which is precisely
what the Hilbert space criterion for Cauchy sequence is. In other
words, even if \textbf{th$_{\text{1}}$}, \textbf{th$_{3}$}, \textbf{th$_{\text{5}}$},
\textbf{th$_{\text{7}}$} are omitted from considerations of \textbf{I};
\textbf{th$_{\text{2}}$}, \textbf{th$_{\text{4}}$} and \textbf{th$_{6}$}
will be parts of \textbf{I}.\textbf{}\\
\textbf{}\\
Hence, the thread-mesh representing any biological threshold level
can be considered as a Hilbert space.\textbf{ }This is supported from
other aspects of thread-mesh also. For example, any thread (a biological
property modeled as function) is a real-valued function and on the
biologically relevant interval of this function, the integral of the
square of its absolute value (over that interval) will necessarily
be finite (no biological property has ever been reported to assume
a non-finite magnitude during its existence). This implies that the
threads (biological properties) can be represented by measurable and
square integrable functions and such a characteristic is a hallmark
of Hilbert space (or putting in other way, \textbf{L$^{\text{2}}$
space}). \\
\\
A careful observation topology of thread-mesh (abstract normed vector
space representing biological properties of any biological threshold
level) reveals several other interesting characteristics of it :\\
\textbf{\underbar{2.5.2.2) Additional topological characteristic of
thread-mesh - 01)}}\\
Geometry of the thread-mesh space resembles that of a real vector
bundle, where we define a vector bundle as a geometric construct which
makes precise the idea of a family of vector spaces parametrized by
another space X. Here X can be a topological space, a manifold, or
a pertinent algebraic construct. The definition demands that if, to
every point x of the space X we can associate a vector space V(x)
in such a way that these vector spaces fit together to form another
space of the same kind as X (e.g. a topological space, manifold, or
a pertinent algebraic construct), it can then be called a vector bundle
over X. Since, the nature of thread-mesh in either 'biological-system'
set or observer space does not differ, we can safely represent the
thread-mesh space by a real vector bundle. Thus it can be said that
:\\
1) V$^{\text{*}}$ (space representing the observer thread-mesh space)
and V (space representing systemic thread-mesh space) are finite-dimensional.\\
2) V$^{\text{*}}$ has the same dimensions as V.\\
Assumption (2) of TM model states that only a subset of the entire
set of observer properties can interact with the 'Biol-System Thread-Mesh'(BSTM).
We denote this subset of observer properties as the 'preference' of
the observer. In a more formal representation, if $(e_{\text{1}},...,e_{\text{n}})$
forms the basis of V (the vector space representing systemic thread-mesh
space); then the associated basis for V$^{\text{*}}$(the vector space
representing observer thread-mesh space) can be written as $(e^{\text{1}},...,e^{\text{n}})$
where :\textbf{}\\
$e^{\text{i}}(e_{\text{j}})=1$ if $i=j$ (observer preferences are
capable of observing BSTM),\\
and\textbf{}\\
$e^{\text{i}}(e_{\text{j}})=0$ if $i\neq j$ (observer preferences
are not capable of observing BSTM).\\
The capability of observer to observe any biological phenomenon stems
from compatibility between thread-set representing observational mechanism
and BSTM. For example, to obtain a measure of structural constraints
of an enzyme, the observer thread-set should necessarily interact
with only certain threads of BSTM (for example, the permitted ranges
of bond lengths and bond angles, omega angle restraints, side chain
planarity, proline puckering, B-factor distribution, rotamer distribution,
Ramachandran plot characteristics etc.). If the observational mechanism
gathers information about the associated pathways, catalytic sites
of the enzyme, its cellular location, or about its functional domains;
they will be incompatible with the systemic properties of interest
and thus will not be capable to obtain a measure of structural constraints
of an enzyme). To elaborate a little, if the systemic thread-mesh
space were a simple 2 dimensional space, \textbf{R$^{\text{2}}$},
its basis B would have been given by : \\
$B={e_{\text{1}}=(1,0),e_{\text{2}}=(0,1)}$. \\
Then, \textbf{e}$^{\text{1}}$ and \textbf{e}$^{\text{2}}$ can be
called one-forms (functions which map a vector to a scalar), \\
such that \textbf{:}\begin{equation}
\mathbf{e^{\text{1}}(e_{\text{1}})=1,e^{\text{1}}(e_{\text{2}})=0,e^{\text{2}}(e_{\text{1}})=0}\mbox{ and }\mathbf{e^{\text{2}}(e_{\text{2}})=1}\end{equation}
\\
\textbf{\underbar{2.5.2.3) Additional topological characteristic of
thread-mesh - 02)}}\\
The BSTM representing any \textbf{Th$_{\text{i}}$}, (say \textbf{X}),
is a topological vector space.\\
Let x and y be the coordinates of any two threads (biological properties)
in \textbf{X}. Although the biological properties are closely dependent
upon one another, they are distinct in their functionality (for example,
for the biological threshold level of proteins, a thread representing
interior mass distribution of a protein may be having a close correlation
with another thread representing interior hydrophobicity distribution
of the same protein; but these threads are indeed different). Thus
in the context of geometry of thread-mesh we can say that x and y
can be separated by neighborhoods if there exists a neighborhood (infinitesimal
or not) N$_{\text{1}}$ of x and N$_{\text{2}}$of y, such that N$_{\text{1}}$
and N$_{\text{2}}$ are disjoint, $\left(N_{\text{1}}\cap N_{\text{2}}=0\right)$.
This property, viz. any two distinct points of \textbf{X} can be separated
by neighborhoods, suggests that the biological thread-mesh (X) under
consideration can be called a \textbf{T$_{\text{2}}$}space (Hausdorff
space).\\
\\
\textbf{\underbar{2.6) Insights into the thread dynamics :}}

If a particular thread denoting property p and coordinate x within
biological space (BSTM, in terms of generalized coordinate), have
a potential V to be part of an interaction; then at any instance of
time t, we can quantify the action A (in Lagrangian formulation as)
:

\begin{eqnarray}
A[x] & = & \int L\left[x(t),\dot{x}(t)\right]dt\\
 & = & \int(\frac{p}{2}\sum\dot{x_{\text{i}}}^{\text{2}}-V(x(t)))dt\nonumber \end{eqnarray}

Here we note that while x (generalized coordinate for any arbitrarily
chosen biological property p of \textbf{TH$_{\text{i}}$}) has an
explicit dependence on time (since biological properties are time-varying);
owing to the symmetry of thread locations in the thread-mesh for any
\textbf{TH$_{\text{i}}$}, V does not. Since x$_{\text{i}}$ stands
for the generalized coordinate of any thread p (biological property;
be it compositional, structural or functional) within any arbitrarily
chosen threshold level \textbf{TH$_{\text{i}}$}, $\dot{x}$ represents
change of position of the thread p; which implies interaction between
threads. More magnitude of $\dot{x}$ implies more interaction between
biological properties.\\
\\
Thus, denoting $Q=\frac{\partial}{\partial t}$ (so that $Q[x(t)]=\dot{x}$
)\\

$Q[L]=p\sum\limits _{\text{i}}\dot{x}\ddot{x}-\sum\limits _{\text{i}}\frac{\partial V(x)}{\partial x_{\text{i}}}\dot{x}=\frac{d}{dt}[\frac{p}{2}\sum\limits _{\text{i}}\dot{x}^{\text{2}}-V(x)]$

If we set

\[
K=[\frac{p}{2}\sum_{\text{i}}\dot{x}^{\text{2}}-V(x)]\]

then

\begin{eqnarray}
z & = & \sum\limits _{\text{i}}\frac{\partial L}{\partial\dot{x}}Q[x_{\text{i}}]-K\nonumber \\
 & = & p\sum_{\text{i}}\dot{x}^{\text{2}}-[\frac{p}{2}\sum_{\text{i}}\dot{x}^{\text{2}}-V(x)]\nonumber \\
 & = & [\frac{p}{2}\sum_{\text{i}}\dot{x}^{\text{2}}+V(x)]\end{eqnarray}

The \textbf{eq$^{\text{n}}$-12}, describes total energy of BSTM for
any arbitrarily chosen \textbf{TH$_{\text{i}}$} and owing to the
presence of symmetrical potential V(x), \textbf{eq$^{\text{n}}$-12}
describes a conservation of z. Since we know that the conservation
of energy is the direct consequence of the translational symmetry
of the quantity conjugate to energy, namely time; an application of
Noether's theorem suggests $\dot{z}=0$. This implies that the principle
of conservation of thread-mesh energy for any \textbf{TH$_{\text{i}}$}
is a consequence of invariance under translation through time.\\
\\
The conformance to Noether's theorem directly suggests that if the
process of observation (over any interval of time), is represented
by an one-dimensional manifold, the systemic thread-mesh space (comprised
of system's properties and pertinent environmental properties) can
be considered as a target manifold of the it. Under such circumstances
one can attach to every point x of a smooth (or differentiable) manifold,
a vector space called the cotangent space at x. Typically, this cotangent
space can be defined as the dual space of the tangent space at x.
Hence the thread-mesh can be thought of as the cotangent bundle of
space of generalized positions of threads, with respect to the observation
manifold.\\
\\
Since cotangent bundle of a smooth manifold can as well be considered
as the vector bundle of all the cotangent spaces at every point in
the manifold, we can assert that observer properties that are compatible
with some systemic property and is capable of measuring it, must be
sharing a canonical relationship with each other. Reasons behind such
argument follow from the special set of attributes that coordinates
on the cotangent bundle of a manifold satisfy. Thus, if 'q's denote
the coordinates on the underlying manifold (systemic thread-mesh space)
and the 'p's denote their conjugate coordinates (observer thread-mesh
space) then they can be written as a set of $(q^{i},p_{j})$ too.
\\
If C denotes the configuration space of smooth functions between thread-mesh
manifold M to the observation manifold O, then the action A (aforementioned)
can be characterized precisely as a functional, $A:C\longrightarrow O$.\\
\\
The biological properties are functions of many influencing factors.
Hence functional A can be precisely described as a function that takes
functions as its argument (from the thread-mesh manifold) and returns
a real number to be perceived by the observer at a given instance
of time (observation manifold).\\
\\
\textbf{\underbar{2.7) Uncertainty in observation of biological phenomenon
:}}\\
The macroscopic observable nature of biological properties come to
existence due to frequency of interactions amongst many biological
properties belonging to biological system or the environment or both.
An individual interaction between two biological entities does not
account for an observable biological property, but significant frequency
of same type of interactions within some defined set of biological
entities do produce a biological property. A series of very recent
findings tend to vindicate this assertion. For example, a significant
frequency of Brownian collisions between parts of protein molecules
(not a single collision) within cytoplasm is what has been suggested
to cause aggregation \cite{Chang2005}; similarly it is found that
the significant frequency (and not a single interaction) with which
the cancer proteins participate in various interactions is what attributes
them their unique nature \cite{Jonsson2006}. The importance of frequency
of biomolecular collisions between macromolecules (representing \textbf{TH$_{\text{biol-macromolecules}}$})
within the cytoplasm, which causes weak emergence at the next threshold
level (viz.\textbf{ TH$_{\text{biochemical pathways}}$}) is discussed
in details by Alsallaq \cite{Alsallaq2007}. From a completely different
paradigm, the significance of frequency of interactions on the ecological
communities have been reported by Tylianakis \cite{Tylianakis2008}.
The effects of all these interactions, viz. measurable biological
properties, are observed in the time domain by the observer. Hence
while the process of observation of any biological property takes
place in the time domain, the macroscopic observable nature of biological
property comes to being due to significant frequency of thread interactions
in the thread-mesh space.\\
Since, first, any arbitrarily chosen biological property can be described
in the relevant biological frequency domain in the thread-mesh space,
when its observation takes place on the the time domain and second,
the thread-mesh manifold has been proved to be residing on the cotangent
bundle of the observation manifold; the entire arrangement can be
described in terms of a Fourier transform pair in the observation
(time) domain. Thus, if any arbitrarily chosen biological property
is observed to be represented as an waveform with basis element b(t)
(that is, in the observation domain) with its Fourier transform $B(\Omega)$
(in the thread-mesh domain); we can define the energy of the waveform
to be E; so that (by Parseval's theorem) :\\

\begin{equation}
E=\int\limits _{-\infty}^{\infty}(|b(t)|)^{2}dt=\frac{1}{2\pi}\int\limits _{-\infty}^{\infty}(|B(\Omega)|)^{2}d\Omega\end{equation}

Since every biological property operates within a specified bound
of magnitude, something that has been referred to as 'fluctuation'
in an earlier study \cite{Testa2000} the functions that represent
them will also be bounded in their ranges. Examples for such fluctuation
are many; in \textbf{TH$_{\text{Cell}}$}, for the mitogen-activated
protein kinase cascade studies, the total concentrations of MKKK,
MKK and MAPK have been found to be in the range 10\textendash{}1000
nm and the estimates for the kcat values of the protein kinases and
phosphatases have been found to range from 0.01 to 1 s$^{\text{-1}}$
\cite{Kholodenko2000}. Similarly, for\textbf{ }the proteins (\textbf{TH$_{\text{biol-macromolecules}}$}),
the mass fractal dimension representing compactness of the protein
has been found to be in the range of 2.22 to 2.69 \cite{Enright2005}.
Thus for any arbitrarily chosen biological property we can identify
a center of the waveform representing the property $t_{\text{c}}$
and $\Omega_{c}$(observed mean magnitude that the property can be
associated to) along with corresponding widths $\mathbf{\Delta_{t}}$
and$\mathbf{\Delta_{\Omega}}$(the permissible limits that the magnitude
of the observed property can approach, $\Delta_{t}^{\text{2}}$ and$\Delta_{\Omega}^{\text{2}}$
can be interpreted as variances of t and $\Omega$). Since biological
properties are represented by threads in the TM model, we can interpret
$t_{c}$ and $\Omega_{c}$as the mean coordinates for the location
of the threads in thread-space, with the variance $\Delta_{t}$ and$\Delta_{\Omega}$
representing their permissible range of variability around $t_{c}$
and $\Omega_{c}$, respectively. Any thread \textbf{Th$_{\text{1}}$}
having more variance than a thread \textbf{Th$_{\text{2}}$}will imply
the biological property represented by the thread \textbf{Th$_{\text{1}}$}is
more capable to interact (and influence) with other biological properties
than the property represented by \textbf{Th$_{\text{2}}$}; or in
other words, the biological property represented by \textbf{Th$_{\text{2}}$}is
more specific in its mode of working than the property represented
by \textbf{Th$_{\text{1}}.$} \\
Hence, \\
1) for central measures of the functions representing biological properties,
we have : 

\begin{equation}
t_{\text{c}}=\frac{1}{E}\int\limits _{-\infty}^{\infty}t(|b(t)|)^{\text{2}}dt\end{equation}

\begin{equation}
\Omega_{\text{c}}=\frac{1}{2\pi E}\int\limits _{-\infty}^{\infty}\Omega(|B(\Omega)|)^{\text{2}}d\Omega\end{equation}

Since both $\frac{1}{E}(|b(t)|)^{\text{2}}$ and $\frac{1}{2\pi E}(|B(\Omega)|)^{\text{2}}$
are non-negative and both of them integrate to 1, they satisfy the
requirements of probability density functions for random variables
t and $\Omega$, with t$_{\text{c}}$ and $\Omega$$_{\text{c}}$denoting
their respective means. Such probabilistic interpretation of the functioning
of biological properties can be immensely helpful in describing any
biological phenomenon, because in many of the cases, a deterministic
knowledge about the extent of involvement of any biological property
in a process is found absent.\\
\\
2) for the deviations (widths) of the functions representing biological
properties, we have : 

\begin{equation}
\Delta_{\text{t}}=\sqrt{\frac{1}{E}\int\limits _{-\infty}^{\infty}\left(t-t_{c}\right)^{2}\left(\left|b(t)\right|\right)^{2}dt}\end{equation}

\begin{equation}
\Delta_{\Omega}=\sqrt{\frac{1}{2\pi E}\int\limits _{-\infty}^{\infty}(\Omega-\Omega_{\text{c}})^{2}(\left|B(\Omega)\right|)^{2}d\Omega}\end{equation}
\\
We can now attempt to prove that the nature of biological reality
as observed by any observer sharing the same biological space-time
as the system under consideration, will always be subjective in nature.
Which follows from the fact that there will always be an uncertainty
in observer's measurement of any biological system that he studies.
This uncertainty will be inherent to the process of observation, because
the observer is a part of the biological space-time that he wishes
to observe; therefore his mere presence is going to disturb the biological
space-time (that includes geometry of thread-mesh) in some definite
manner.\\
Hence a super-observer, who is not a part of biological space-time,
will be able to notice that the more an observer (everyone of us)
attempts to locate a thread-mesh property under observation (say,
signal) in the time-domain for a precise measurement; the less would
he able to locate it's nature in the systemic thread-mesh domain.
Because for a very short duration of observation, the observer can
only, at best, hope to capture a mere snap-shot of the thread dependencies
and a snap-shot of causality behind thread interaction behind the
observed biological phenomenon. Although this snap-shot of these dependencies
can be obtained in a precise manner, it will not be able to provide
any insight about either the cause of these dependencies (between
biological properties) or evolution of the dependencies (between biological
properties). Therefore an attempt to locate a signal in the time-domain
(preciseness in the observation manifold) will result in obtaining
an inherently incomplete and inadequate description of the biological
process under consideration. On the other hand, if the observation
process is carried out over a long duration of time, only the statistical
nature of thread interactions (statistical nature of biological properties)
can be measured and not the precise causalities and time-variant dependencies
behind thread interactions. Thus even in this case also, only an incomplete
idea of biological reality can be found with uncertainty about the
precise biological causes and time-variant precise dependencies (between
biological properties) behind the process.\\
\\
\textbf{\underbar{2.8) Uncertainty relationship for Biology :}}\\
The canonical conjugacy between variables chosen from thread-mesh
space and observation manifold can be expressed in the form of an
uncertainty principle for biology. This fundamental uncertainty in
observation of any biological property can be mathematically expressed
as :

\textbf{\underbar{Theorem-2 :}} \textbf{If $\sqrt{|t|}b(t)\rightarrow0\mbox{ as }|t|\rightarrow\infty$
then}

\textbf{$t_{\text{c}}\Omega_{\text{c}}\geq\frac{1}{2}$}

\textbf{and the equality holds only if b(t) is of the form $b(t)=Ce^{\text{-}\alpha t^{\text{2}}}$
}\\
\textbf{\underbar{Proof :}} The Cauchy-Schwarz inequality for any
square integrable functions z(x) and w(x) defined on the interval
{[}a,b] states, 

\begin{equation}
\left|\intop_{a}^{b}z(x)w(x)dx\right|^{2}\leq\intop_{a}^{b}\left|z(x)\right|^{2}dx\intop_{a}^{b}\left|w(x)\right|^{2}dx\end{equation}

Since b(t) is real for the biological properties, an application of
last equation yields\\
\begin{equation}
\left|\intop\limits _{-\infty}^{\infty}tb\frac{db}{dt}dt\right|^{2}\leq\intop\limits _{-\infty}^{\infty}t^{\text{2}}b^{\text{2}}dt\intop\limits _{-\infty}^{\infty}|\frac{db}{dt}|^{\text{2}}dt\end{equation}

let \begin{eqnarray*}
A & = & \intop_{-\infty}^{\infty}tb\frac{db}{dt}dt\\
 & = & \int t\frac{d(b^{2}/2)}{dt}dt\\
 & = & \underset{\alpha}{\underbrace{t\frac{b^{2}}{2}|_{-\infty}^{\infty}}}-\underset{\beta}{\underbrace{\intop_{-\infty}^{\infty}\frac{b^{2}}{2}dt}}\end{eqnarray*}
In the limit, $\sqrt{|t|}b\rightarrow0\Rightarrow|t|b^{\text{2}}\rightarrow0\Rightarrow tb^{\text{2}}=0.$
Thus $\alpha=0.$ \\
Furthermore $\beta=E/2$ (from \textbf{eq$^{\text{n}}$-13}) ; and
so 

\begin{equation}
A=-E/2\end{equation}

Recalling that$\frac{d}{dt}b(t)\leftrightarrow\mathit{j\Omega}B(\Omega)$,
by Parseval's theorem we have :

\begin{equation}
\int\limits _{-\infty}^{\infty}|\frac{db}{dt}|^{\text{2}}dt=\frac{1}{2\pi}\int\limits _{-\infty}^{\infty}\Omega^{\text{2}}|B(\Omega)|d\Omega\end{equation}

Substituting (\textbf{eq$^{\text{n}}$-20}) and (\textbf{eq$^{\text{n}}$-21})
into (\textbf{eq$^{\text{n}}$-19}) we obtain :\\
\begin{eqnarray}
|-\frac{E}{2}|^{\text{2 }} & = & |\int\limits _{-\infty}^{\infty}tb\frac{db}{dt}dt|^{\text{2}}\\
 &  & \leq\underset{Et^{2}}{\underbrace{\int\limits _{-\infty}^{\infty}t^{\text{2}}b^{\text{2}}dt}}\times\underset{E\Omega^{2}}{\underbrace{\frac{1}{2\pi}\int\limits _{-\infty}^{\infty}\Omega^{\text{2}}(|B(\Omega)|)^{\text{2}}d\Omega}}\end{eqnarray}
\\
\begin{equation}
\Rightarrow t_{c}\Omega_{c}\geq\frac{1}{2}\end{equation}
\\
In the special case, if (\textbf{eq$^{\text{n}}$-24}) is an equality,
then (\textbf{eq$^{\text{n}}$}-\textbf{19}) must be also; which is
possible only if

$\frac{d}{dt}$b(t) = k t b(t)

$\Rightarrow$b(t) = Ce$^{\text{-}\alpha t^{\text{2}}}$ (a Gaussian
waveform). \textbf{\hfill{}Q.E.D}\\

\section*{3.\underbar{ Experimental works that report }}

\section*{\underbar{observer-dependence in understanding biological reality
:}}

The inherent constraint of observer dependence in knowing biological
reality, as have been proved (\textbf{eq$^{\text{n}}$-24}) with the
TM model, echoes (mathematically) the findings of many previous studies.
The prophetic views of Ashby \cite{Ashby1973} had stressed on the
importance of acknowledging observer-dependence, so did Kay \cite{Kay1984}
in his assessment of scope of application of information theory to
biological systems, and so did a list of works who had touched upon
the role of observer-dependence in the studies of emergence and complexity
\cite{Casti1986,Cariani1991,Baas1994,Brandts1997}. To consider particular
examples, how the studies on p53 gene mutation and/or p53 protein
expression can be observer-dependent (owing to the innate nature of
immunohistochemical techniques \cite{Feilchenfeldt2003}) and how
therefore, they report different views of biological reality, have
been documented in a review \cite{Ishii1998}. From a completely different
paradigm of studies involving cerebral cortex, the cytoarchitectonical
distinctions are also reported to suffer from observer-dependence,
resulting in several definitions of cortical areas \cite{Kotter2001}.
From using immunocytology as an observer-dependent standard method
for tumour cell detection \cite{Benoy2004}, to observer-dependent
techniques involving immunocytochemistry in attempting to quantify
neurodegeneration in animal models \cite{Petzold2003}; from procedures
of cell-counting in epidemiological studies \cite{Araujo2004}, to
ways of noninvasive assessments of endothelial function that usually
relies on postischaemic dilation of forearm vessels\textbf{ }and use
of flow-mediated dilation measurements of brachial artery \cite{Lee2002};
from aspects of immunofluorescence testing \cite{Meda2008} to methods
involving positron emission tomography in the realm of radiotherapy
\cite{Jarritt2006} - observer-dependence in ascertaining the biological
reality is well documented in myriad contexts. The diversity of biological
realms that report observer-dependence, tend to point to the universal
presence of it in our (observer's) attempts to know biological reality.
From the case of so called {}``bystander effect'', where nanoparticle
mediated cell transfection study was reported to suffer from {}``observer
effect'' \cite{Zhang2007}; to studies involving tumor peripheries
in the context of breast tumors \cite{Preda2005}; experimental biology
is replete with reports of observer dependence in understanding biological
reality. \\
\\
Some of the most startling studies in this regard prove unmistakably
that observer's mere presence perturbs the biological reality. This
happens because conducting an experiment on some system residing within
biological space-time corresponds to an active perturbation of the
thread-mesh by the observer. However, since the observer is always
a part of the thread-mesh (regardless of his being an experimentor
or not), his sheer presence is going to perturb the thread-mesh geometry
in some passively subtle, yet definite manner. The existence of the
observer in the biological space-time implies that he (observer) extracts
necessary energy for his survival from the same pool of available
energy, which the biological system under consideration is also using
to derive energy from. Hence this act of sharing the available (solar)
energy between the biological system and the observer ensures that
the presence of observer perturbs the biological reality, across all
threshold levels. This point is proved in a recent experimental work
\cite{DeBoeck2008} where the authors refute the notion of possible
existence of any {}``benign observer''. The perturbation of biological
reality by the very existence of the observer has been strongly reported
in many other experimental works too \cite{Almeida2006,Siegfried2006,Lay1999,Wahlsten2003,Hik2003}.
Even the possibility of an uncertainty relationship arising out of
observer's interference with biological reality in the realm of ecology
was discussed in systematic well-documented manner by Cahill \cite{Cahill2001}
and (in philosophical terms) in another work \cite{Regan2002}. However
the uncertainty relation (\textbf{eq$^{\text{n}}$-24}) is significant
because it mathematically proves that for any super-observer (who
is not a part of biological space-time) it will be clear that during
any observation process (passive/active), an attempt to measure any
biological property exactly(typically in the time-domain, because
evolution of biological properties are observed over relevant time-scales)
will not be able to capture the complete information of the same in
biologically functional space. This is because precise measurements
will only capture a snap-shot of the property-dependencies prevalent
in the system under consideration; whereas a biological meaningful
complete description of these time-dependent and context-dependent
property-dependencies can only be provided if the measurement is conducted
over a long period of time; which on the other hand, due to its innate
statistical nature, smoothen out the finer aspects of phenomenon under
consideration. Hence, it is in the very nature of biological reality
that it would not be observed in an objective way.

\section*{4.\underbar{ Conclusion :}}

A description-oriented theoretical toy model to study the nature of
biological reality, the \textquoteleft{}thread-mesh model\textquoteright{},
has been proposed. The necessity to construct this model originated
from the realization that issues related to biological complexity
can honestly be answered only when the nature of biological reality
can be objectively described, qualitatively and quantitatively. The
TM model attempts to mimic biological reality by, first, splitting
the entire biological universe into a series of biological threshold
levels (different biological organization) and second; by describing
the compositional, structural and functional features of these biological
threshold levels (with their specific environmental constraints).
The proposed model used a linear algebraic framework to describe biological
complexity acknowledging emergence. It could describe and explain
the context-dependent nature of biological behaviors. The role of
observer in measuring a biological property is exhaustively examined
in the proposed thread-mesh paradigm and it has been proved here that
nature of biological reality is observer dependent. Taking a note
of canonical conjugacy between variables chosen from biological property
manifold and observation manifold, an uncertainty relationship for
biology has been suggested too, which proved that there cannot be
an objective description of biological reality and therefore an objective
and complete complexity measure can not exist for biological systems.
The nature of biological complexity measures can only be subjective
(to varying extent) and can only be relevant when the scope of them
(biological threshold level) is mentioned.\\
\textbf{}\\
\textbf{Acknowledgment} This work was supported by COE-DBT (Department
of Biotechnology, Government of India) scholarship.\\
The author wants to thank professor Indira Ghosh, former Director,
Bioinformatics centre, University of Pune along with Dr. Urmila Kulkarni-Kale,
the present Director, Bioinformatics centre, University of Pune for
providing him with an opportunity to work on a problem that has got
nothing to do with his PhD degree.


\begin{thebibliography}{Dokholyan and Shakhnovich, 2001}
\bibitem[Abel and Trevor, 2005]{Abel2005}Abel, D.L, Trevors, J.T.,
2005. Three subsets of sequence complexity and their relevance to
biopolymeric information. Theor. Biol. and Med. Modelling. 2, 29.

\bibitem[Adami, 2002]{Adami2002}Adami, C., 2002. What is complexity?
BioEssays. 24, 1085\textendash{}1094.

\bibitem[Alba et al., 2002]{Alba2002}Alba, M.M., Laskowski, R.A.,
Hancock, J.M., 2002. Detecting cryptically simple protein sequences
using the SIMPLE algorithm. Bioinformatics.18, 672-678.

\bibitem[Almeida et al., 2006]{Almeida2006}Almeida, M., Paula, H.,
Tavora, R., 2006. Observer effects on the behavior of non-habituated
wild living Marmosets (Callithrix jacchus); Revista de Etologia. 8(2),
81-87.

\bibitem[Alsallaq and Zhou, 2007]{Alsallaq2007}Alsallaq, R., Zhou,
H.X., 2007, Energy Landscape and Transition State of Protein-Protein
Association, Biophys. J., 92, 1486-1502.

\bibitem[Anand and Tucker, 2003]{Anand2003}Anand, M., Tucker, B.C.,
2003. Defining biocomplexity: an ecological perspective; Comm. Theor.
Biol. 8, 497-510.

\bibitem[Andrianantoandro et al., 2006]{Andrian2006}Andrianantoandro,
E., Basu, S., Karig, D.K., Weiss, R., 2006. Synthetic biology: new
engineering rules for an emerging discipline; Mol. Syst. Biol. 2,
0028.

\bibitem[Aon et al., 2004]{Aon2004}Aon, M.A., O'Rourke, B., Cortassa,
S., 2004. The fractal architecture of cytoplasmic organization: Scaling,
kinetics and emergence in metabolic networks. Mol. Cell. Biochem.
256-257: 169\textendash{}184.

\bibitem[Araujo et al., 2004]{Araujo2004}Araujo, R., Rodrigues, A.G.,
Pina-Vaz, C., 2004. A fast, practical and reproducible procedure for
the standardization of the cell density of an Aspergillus suspension.
J. Med. Microbiol. 53, 783-786.

\bibitem[Ashby, 1973]{Ashby1973}Ashby, W., 1973. Some peculiarities
of complex systems. Cybernetics Med. 9, 1-8.

\bibitem[Baas, 1994]{Baas1994}Baas, N., 1994. Emergence, Hierarchies,
and Hyperstructures. In: Langton C. G., (Ed.) Art. life III, Santa
Fe Inst. Studies in the Sciences of Complexity Proc. Redwood City,
Calif. Addison-Wesley. Vol. XVII., pp.515-537.

\bibitem[Badii and Politi, 1997]{Badii1997}Badii, R., Politi, A.,
1997. Complexity: Hierarchical Structures and Scaling in Physics;
Cambridge University Press, Cambridge, UK.

\bibitem[Banerji and Yeragani, 2003]{Banerji2003}Banerji, A., Yeragani,
V.K., 2003. A new measure to quantify the complexity of phase data
(PhaseCmp) from cross-spectral analysis. Cardiovascular Eng. 3(4),
149-154.

\bibitem[Barberis et al., 2007]{Barberis2007}Barberis, M., Klipp,
E., Vanoni, M., Alberghina, L., 2007. Cell Size at S Phase Initiation:
An Emergent Property of the G1/S Network. PLoS Comp. Biol. 3(4), e64.

\bibitem[Barcellos-Hoff, 2008]{Barcellos2008}Barcellos-Hoff, M.,
2008. Cancer as an emergent phenomenon in systems radiation biology;
Radiat. Environ. Biophys. 47(1), 33-38.

\bibitem[Barnstad et al., 2001]{Barnstad2001}Bjørnstad, O.N., Sait
S.M., Stenseth, N.C., Thompson, D.J., Begon, M., 2001. The impact
of specialized enemies on the dimensionality of host dynamics. Nature.
409, 1001-1006.

\bibitem[Bar-Yam, 1997]{BarYam1997}Bar-Yam., Y., 1997. Dynamics of
Complex Systems, chapter 0, Overview: The Dynamics of Complex Systems
- examples, questions, methods and concepts. Studies in Nonlinearity.
Westview Press.

\bibitem[Bar-Yam, 2004]{Bar-Yam2004}Bar-Yam, Y., 2004. Multiscale
complexity/entropy; Adv. in Complex Sys. 7(1), 47\textendash{}63.

\bibitem[Bedau, 1997]{Bedau1997}Bedau, M., 1997. Weak emergence.
Phil. Perspectives. 11, 375\textendash{}399.

\bibitem[Bennett, 1988]{Bennett1988}Bennett, C., 1988. Logical depth
and physical complexity. In: Herken R. (Ed.), Universal Turing Machine,
A Half-Century Survey, Oxford University Press, Oxford, pp 227\textendash{}257.

\bibitem[Benoy et al., 2004]{Benoy2004}Benoy, I.H., Elst, H., Van
der Auwera, I., Van Laere, S., van Dam P., Van Marck, E.,Scharpé,
S., Vermeulen P.B., Dirix L.Y., 2004. Real-time RT\textendash{}PCR
correlates with immunocytochemistry for the detection of disseminated
epithelial cells in bone marrow aspirates of patients with breast
cancer; Brit. J. Cancer. 91, 1813\textendash{}1820.

\bibitem[Besedovsky et al., 1979]{Besedovsky1979}Besedovsky, H.O.,
del Rey, A., Sorkin, E., Da Prada, M., Keller, H.H., 1979. Immunoregulation
mediated by the sympathetic nervous system. Cell Immunol., 48(2),
346-55.

\bibitem[Bialek et al., 2001]{Bialek2001}Bialek, W., Nemenman, I.,
Tishby, N., 2001. Predictability, Complexity, and Learning; Neural
Comp. 13, 2409-2463.

\bibitem[Bonabeau and Dessalles, 1997]{Bonabeau1997}Bonabeau, E.,
Dessalles, J.L., 1997. Detection and emergence. Intellectica. 25,
89-94.

\bibitem[Bonchev, 2003]{Bonchev2003}Bonchev D., 2003. On the complexity
of directed biological networks. SAR and QSAR in Env. Res. 14(3),
199-214.

\bibitem[Boogerd et al.(2005)]{Boogerd2005}Boogerd, F., Bruggeman,
F., Richardson, R., Stephan, A., Westerhoff, H., 2005. Emergence and
its place in nature: a case study of biochemical networks. Synthese.
145, 131-164.

\bibitem[Brandts, 1997]{Brandts1997}Brandts, W.A.M., 1997. Complexity:
a pluralistic approach. In: Lumsden, C.J., Brandts, W.A., Trainor,
L.E.H. (Eds.) Physical theory in biology. Foundations and explorations.
Singapore: World Scientific. pp. 45-68.

\bibitem[Cahill et al., 2001]{Cahill2001}Cahill, J.F.Jr., Castelli,
J.P., Casper, B.B., 2001. The herbivory uncertainty principle: visiting
plants can alter herbivory. Ecology. 82(2), 307-312.

\bibitem[Cariani, 1991]{Cariani1991}Cariani, P., 1991. Emergence
and Artificial Life. In: Langton C. G., Taylor C., Farmer J. D., Rasmussen
S., (Eds.) Art. Life II. Santa Fe Inst. Studies in the Sciences of
Complexity Proc. Redwood City, Calif. Addison-Wesley. Vol.X, pp.775-797.

\bibitem[Carothers et al., 2004]{Carothers2004}Carothers, J.M., Oestreich,
S.C., Davis, J.H., Szostak, J.W., 2004. Informational complexity and
functional activity of RNA structures; J. Am. Chem. Soc. 126, 5130-5137.

\bibitem[Casti, 1986]{Casti1986}Casti, J., 1986. On system complexity:
identification, measurement, and management. In: Casti J. L., Karlqvist
A., (Eds.), Complexity, Language, and Life: mathematical approaches
(Biomathematics vol.16). Berlin: Springer-Verlag. pp. 146-173.

\bibitem[Chaitin, 1966]{Chaitin1966}Chaitin, G.J., 1966. On the length
of programs for computing finite binary sequences. Journal of the
ACM (JACM). 13, 547-569.

\bibitem[Chang et al., 2005]{Chang2005}Chang, C., Lina P., Yeh X.,
Deng K., Ho Y., Kan L., 2005. Protein folding stabilizing time measurement:
A direct folding process and three-dimensional random walk simulation.
Biochem. Biophys. Res. Comm. 328, 845-850.

\bibitem[Chen et al., 1999]{Chen1999}Chen, X., Kwong, S., Li, M.,
1999. Compression algorithm for DNA sequences and its applications
in genome comparison. Genome Inform. Ser. Workshop Genome Inform.
10, 51-61.

\bibitem[Chiappini et al., 2005]{Chiappini2005}Chiappini E., Galli
L., Tovo P., Gabiano C., Martino M., Thcantatae Itacantatalian Register
for HIV Infection in Children, 2005. Persistently high IgA serum levels
are a marker of immunological or virological failure of combined antiretroviral
therapy in children with perinatal HIV-1 infection. Clin Exp Immunol.
140(2), 320\textendash{}324.

\bibitem[Chuzhanova et al., 2000]{Chuzhanova2000}Chuzhanova, N.A.,
Anassis, E.J., Ball, E.V., Krawczak, M., Cooper, D.N., 2000. Meta-analysis
of indels causing human genetic disease: mechanisms of mutagenesis
and the role of local DNA sequence complexity. Hum. Mutation. 21,
28-44.

\bibitem[Cohen and Atlan, 2006]{Cohen2006}Cohen, I., Atlan, H., 2006.
Genetics as explanation: Limits to the human genome project. Encyclopedia
of Life Sci., 2006, John Wiley \& Sons, Ltd., 1-7.

\bibitem[Cornacchio, 1977]{Cornacchio1977}Cornacchio, J., 1977. Maximum-entropy
complexity measures, Int. J. Gen. Sys., 3, 215\textendash{}225.

\bibitem[Cox and Mirkin, 1997]{Cox1997}Cox, R., Mirkin, S.M., 1997.
Characteristic enrichment of DNA repeats in different genomes. Proc.
Natl. Acad. Sci. USA. 94, 5237-5242.

\bibitem[Crutchfield, 1994]{Crutchfield1994}Crutchfield, J., 1994.
Is anything ever new? Considering emergence. Working paper, Santa
Fe Institute. 94-03-011.

\bibitem[Crux et al., 1997]{Crux1997}Cruz, X., Mahoney, M., Lee,
B., 1997. Discrete representations of the protein C$\alpha$ chain.
Folding and Design. 2, 223\textendash{}234.

\bibitem[Davies, 2004]{Davies2004}Davies, P.C.W., 2004. Emergent
biological principles and the computational properties of the universe.
Complexity. 10(2), 11-15.

\bibitem[De Boeck et al., 2008]{DeBoeck2008}De Boeck, H., Liberloo,
M., Gielen, B., Nijs, I., Ceulemans, R., 2008. The observer effect
in plant science. New Phytologist, 177, 579\textendash{}583.

\bibitem[Dhar, 2007]{Dhar2007}Dhar, P., 2007. The next step in biology:
A periodic table? J. Biosci. 32, 1005\textendash{}1008.

\bibitem[Ding and Dokholyan, 2006]{Ding2006}Ding, F., Dokholyan,
N.V., 2006. Emergence of protein fold families through rational design.
PloS Comp. Biol. 2, 725-733.

\bibitem[Dokholyan and Shakhnovich, 2001]{Dokholyan2001}Dokholyan,
N.V., Shakhnovich, E.I., 2001. Understanding hierarchical protein
evolution from first principles. J. Mol. Biol. 312, 289-307.

\bibitem[Ebeling et al., 2001]{Ebeling2001}Ebeling, W., Steuer, R.,
Titchener, M.R., 2001. Partition-based entropies of deterministic
and stochastic maps. Stochastics and Dyn. 1(1), 1\textendash{}17.

\bibitem[Edmonds, 1999]{Edmonds1999}Edmonds, B., 1999. Syntactic
measures of complexity. PhD thesis, University of Manchester.

\bibitem[Enright and Leitner, 2005]{Enright2005}Enright, M.B., Leitner,
D.M., 2005. Mass fractal dimension and the compactness of proteins.
Phys. Rev. E., 71, 011912.

\bibitem[Feilchenfeldt et al., 2003]{Feilchenfeldt2003}Feilchenfeldt,
J., Tötsch, M., Sheu, S.Y., Robert, J., Spiliopoulos, A., Frilling,
A., Schmid, K.W., Meier, C.A., 2003. Expression of galectin-3 in normal
and malignant thyroid tissue by quantitative PCR and immunohistochemistry.
Mod. Pathol. 16(11), 1117\textendash{}1123.

\bibitem[Fei and Adjeroh, 2004]{Fei2004}Fei, N., Adjeroh, D.A., 2004.
On complexity measures for biological sequences. Computational Systems
Bioinformatics Conference, 2004. CSB 2004. Proceedings. 2004 IEEE.
522-526.

\bibitem[Fluckiger, 1995]{Fluckiger1995}Flückiger, D.F., 1995. Contributions
towards a unified concept of information. PhD thesis; University of
Berne.

\bibitem[Furusawa et al., 1995]{Furusawa1998}Furusawa, C., Kaneko,
K., 1998. Emergence of rules in cell society: differentiation, hierarchy
and stability. Bull. Math. Biol. 60(4), 659-87.

\bibitem[Gabrielian and Bolshoy, 1999]{Gabrielian1999}Gabrielian,
A.E., Bolshoy, A., 1999 . Sequence complexity and DNA curvature. Comput.
Chem. 23, 263-274.

\bibitem[Gellmann, 1994]{Gellmann1994}Gell-mann, M., 1994. The Quark
and the Jaguar (W.H. Freeman and Co.).

\bibitem[Gellmann and Lloyd, 1996]{Gellmann1996}Gell-mann, M., Lloyd,
S., 1996. Information measures, effective complexity, and total information.
Complexity. 1(1), 44\textendash{}52.

\bibitem[Glade et al., 2004]{Glade2004}Glade, N., Demongeot, J.,
Tabony, J., 2004. Microtubule self-organisation by reaction-diffusion
processes causes collective transport and organisation of cellular
particles. BMC Cell Biol. 5, 23.

\bibitem[Grassberger, 1986]{Grassberger1986}Grassberger P., 1986.
Toward a quantitative theory of self\textendash{}generated complexity.
Int. J. Theor. Phys. 25, 907\textendash{}938.

\bibitem[Gusev et al., 1999]{Gusev1999}Gusev, V.D., Nemytikova, L.A.,
Chuzhanova, N.A., 1999. On the complexity measures of genetic sequences.
Bioinformatics. 15, 994-999.

\bibitem[Hall et al., 1982]{Hall1982}Hall, N.R., McClure, J.E., Hu,
S.K., Tare, N.S., Seals, C.M., Goldstein, A.L., 1982. Effects of 6-hydroxydopamine
upon primary and secondary thymus dependent immune responses. Immunopharmacology.
5(1), 39\textendash{}48.

\bibitem[Hancock, 2002]{Hancock2002}Hancock, J.M., 2002. Genome size
and the accumulation of simple sequence repeats: implications of new
data from genome sequencing projects. Genetica. 115, 93-103.

\bibitem[Hazen et al., 2007]{Hazen2007}Hazen, R.M., Griffin, P.L.,
Carothers, J.M., Szostak, J.W., 2007. Functional information and the
emergence of biocomplexity. Proc. Natl. Acad. Sc. USA, 104, 8574\textendash{}8581.

\bibitem[Hik et al., 2003]{Hik2003}Hik, D., Brown, M., Dabros, A.,
Weir, J., Cahill, J.Jr., 2003. Prevalence and predictability of handling
effects in field studies : results from field experiments and a meta-analysis.
Am. J. Botany. 90(2), 270\textendash{}277.

\bibitem[Hildegard, 2003]{Hildegard2003}Hildegard, M., 2003. Functional
complexity measure for networks; arXiv:cond-mat/0311109 

\bibitem[Hinegardner and Engelberg, 1983]{Hinegardner1983}Hinegardner,
R., Engelberg, H., 1983. Biological complexity. J. Theor. Biol. 104,
7-20.

\bibitem[Hulata et al., 2005]{Hulata2005}Hulata, E., Volman, V.,
Ben-Jacob, E., 2005. Self-regulated complexity in neural networks.
Natural Comp. 4, 363\textendash{}386.

\bibitem[Hut et al., 2000]{Hut2000}Hut, P., Goodwin, B., Kauffman,
S., 2000. Complexity and functionality: a search for the where, the
when, and the how. Proc. Int. Conf. on Complex Sys. Nashua. New Hampshire,
USA. 259-268.

\bibitem[Ishii and Tribolet, 1998]{Ishii1998}Ishii, N., Tribolet,
N., 1998. Are p53 mutations and p53 overexpression prognostic factors
for astrocytic tumors? Crit. Rev. Neurosurg. 8, 269\textendash{}274.

\bibitem[Ishimura et al., 2000]{Ishimura2000}Ishimura, A., Maeda,
R., Takeda, M., Kikkawa, M., Daar, I.O., Maéno. M., 2000. Involvement
of BMP-4/msx-1 and FGF pathways in neural induction in the Xenopus
embryo. Dev. Growth Differ. 42, 307\textendash{}316.

\bibitem[Rangamani and Iyengar, 2007]{Iyengar2007}Rangamani, P.,
Iyengar, R., 2007. Modelling spatio-temporal interactions within the
cell. J. Biosci. Ind. Acad. Sci. 32(1), 157\textendash{}167. 

\bibitem[Jarritt et al., 2006]{Jarritt2006}Jarritt, P.H., Carson,
K.J., Hounsell, A.R., Visvikis, D., 2006. The role of PET/CT scanning
in radiotherapy planning. The Brit. J. Radiol. 79, S27\textendash{}S35.

\bibitem[Jimenez et al., 2002]{Jimenez2002}Jiménez-Montaño, M.A.,
Ebeling, W., Pohl, T., Rapp, P.E., 2002. Entropy and complexity of
finite sequences as fluctuating quantities. Biosystems. 64, 23-32.

\bibitem[Jonsson et al., 2006]{Jonsson2006}Jonsson, P.F., Bates P.A,
2006. Global topological features of cancer proteins in the human
interactome; Bioinformatics. 22, 2291\textendash{}2297.

\bibitem[Kaneko, 1998]{Kaneko1998}Kaneko, K., 1998. Life as complex
systems: Viewpoint from intra-inter dynamics. Complexity. 3(6), 53-60.

\bibitem[Kay, 1984]{Kay1984}Kay, J., 1984. Self-organization in living
systems. Chap.3, PhD thesis. University of Waterloo.

\bibitem[Kholodenko, 2000]{Kholodenko2000}Kholodenko, B., 2000. Negative
feedback and ultrasensitivity can bring about oscillations in the
mitogen-activated protein kinase cascades. Eur. J. Biochem. 267(6),
1583\textendash{}1588.

\bibitem[Kohm and Sanders, 1999]{Kohm1999}Kohm, A.P., Sanders, V.M.,
1999. Suppression of antigen-specific Th2 cell-dependent IgM and IgG1
production following norepinephrine depletion in vivo. J. Immunol.
162, 5299\textendash{}5308.

\bibitem[Kolmogorov, 1965]{Kolmogorov1965}Kolmogorov, A., 1965. Three
approaches to the quantitative definition of information. Problems
of Inf. Transmission. 1, 1-17.

\bibitem[Kotter et al., 2001]{Kotter2001}Kötter, R., Stephan, K.E.,
Palomero-Gallagher, N., Geyer, S., Schleicher, A., Zilles, K., 2001.
Multimodal characterisation of cortical areas by multivariate analyses
of receptor binding and connectivity data. Anat. Embryol. 204, 333\textendash{}350.

\bibitem[Kruszewska et al., 1995]{Kruszewska1995}Kruszewska, B.,
Felten, S., Moynihan, J.A., 1995. Alterations in cytokine and antibody
production following chemical sympathectomy in two strains of mice.
J. Immunol. 155(10). 4613-4620.

\bibitem[Landauer, 1967]{Landauer1967}Landauer, R., 1967. Wanted:
a physically possible theory of physics; IEEE Spectrum. 4, 105\textendash{}109.

\bibitem[Laughlin and Pines, 2000]{Laughlin2000}Laughlin, R., Pines,
D., 2000. The Theory of Everything. Proc. Natl. Acad. Sci. USA. 97,
28-31.

\bibitem[Lay et al., 1999]{Lay1999}Lay, D.C. Jr., Buchanan, H.S.,
Haussmann, M.F., 1999. A note on simulating the \textquoteleft{}observer
effect' using constant photoperiod on nursery pigs. Appl. Animal Behav.
Sc. 63(4), 301-309. doi:10.1016/S0168-1591(99)00018-0

\bibitem[Lee et al., 2002]{Lee2002}Lee, K.W, Felmeden, D.C, Lip,
G.Y., 2002. Statins and the assessment of endothelial function. J.
Internal Med. 251, 452\textendash{}454.

\bibitem[Lempel and Ziv, 1976]{Lempel1976}Lempel, A., Ziv, J., 1976.
On the complexity of finite sequences. IEEE Trans. Inf. Theory. 22,
75-81.

\bibitem[Levchenko et al., 1997]{Levchenko1997}Levchenko, I., Smith,
C.K., Walsh, N.P., Sauer, R.T., Baker, T.A., 1997. PDZ-like domains
mediate binding specificity in the Clp/Hsp100 family of chaperones
and protease regulatory subunits. Cell. 91, 939\textendash{}947.

\bibitem[Li et al., 1996]{Li1996}Li H., Helling R., Tang C., Wingreen
N., 1996. Emergence of preferred structures in a simple model of protein
folding. Science. 273(5275), 666-669.

\bibitem[Lijnzaad and Argos,1997]{Argos1997} Lijnzaad P., Argos P.,
1997. Hydrophobic patches on protein subunit interfaces: characteristics
and prediction. Proteins: Struc. Func. Genet. 28, 333\textendash{}343.

\bibitem[Livnat et al., 1985]{Livnat1985}Livnat, S., Felten, S.Y.,
Carlson, S.L., Bellinger, D.L., Felten, D.L., 1985. Involvement of
peripheral and central catecholamine systems in neural-immune interactions.
J. Neuroimmunol. 10(1), 5-30.

\bibitem[Lofgren, 1977]{Lofgren1977}Lofgren, L., 1977. Complexity
of descriptions of systems: a foundational study. Int. J. General
Sys. 3, 197\textendash{}214.

\bibitem[Madden et al., 1989]{Madden1989}Madden, K., Felten, S.,
Felten, D., Sundaresan, P., Livnat, S., 1989. Sympathetic neural modulation
of the immune system. I. Depression of T cell immunity in vivo and
in vitro following chemical sympathectomy. Brain Behav. Immun. 3,
72-89.

\bibitem[Mahner and Kary, 1997]{Mahner1997}Mahner, M., Kary, M.,
1997. What exactly are genomes, genotypes and phenotypes? And what
about phenomes? J. Theor. Biol. 186, 55-63.doi:10.1006/jtbi.1996.0335

\bibitem[Marguet et al., 2007]{Marguet2007}Marguet, P., Balagadde,
F., Tan, C., You, L., 2007. Biology by design: reduction and synthesis
of cellular components and behaviour. J. R. Soc. Interface. 22, 4(15),
607-623.

\bibitem[Meda et al., 2008]{Meda2008}Meda, F., Zuin, M., Invernizzi,
P., Vergani, D., Selmi, C., 2008. Serum autoantibodies: A road map
for the clinical hepatologist. Autoimmunity, 41(1), 27\textendash{}34.

\bibitem[Miles et al., 1981]{Miles1981}Miles, K., Quintáns, J., Chelmicka-Schorr,
E., Arnason, B.G., 1981. The sympathetic nervous system modulates
antibody response to thymus-independent antigens. J. Neuroimmunol.
1, 101-105.

\bibitem[Milosavljevic and Jurka, 1993]{Milosavljevic1993}Milosavljevic,
A., Jurka, J., 1993. Discovering simple DNA sequences by the algorithmic
significance method. Comput. Appl. Biosci. 9(4),407-411.

\bibitem[Moen et al., 2005]{Moen2005}Moen, D., Winne, C., Reed, R.,
2005. Habitat-mediated shifts and plasticity in the evaporative water
loss rates of two congeneric pit vipers (Squamata, Viperidae, Agkistrodon).
Evol. Ecol. Res. 7, 759\textendash{}766.

\bibitem[Neuwald et al., 1999]{Neuwald1999}Neuwald, A.F., Aravind,
L., Spouge, J.L., Koonin, E.V., 1999. AAA+: A Class of Chaperone-Like
ATPases Associated with the Assembly, Operation, and Disassembly of
Protein Complexes. Genome Res. 9,27\textendash{}43.

\bibitem[Nour et al., 1994]{Nour1994}Nour, S., Fernandez, M., Normand,
P., Clayet-Marel, J., 1994. Rhizobium ciceri sp. nov., consisting
strains that nodulate chickpeas (Cicer artenium L.). Int. J. Syst.
Bacteriol. 44, 511\textendash{}522.

\bibitem[Odell, 2002]{Odell2002}Odell, J., 2002. Agents and Complex
Systems. J. Obj. Tech. 1, 35\textendash{}45.

\bibitem[Orlov et al., 2002]{Orlov2002}Orlov, Y.L., Filippov, V.P.,
Potapov, V.N., Kolchanov, N.A., 2002. Construction of stochastic context
trees for genetic texts. In Silico Biology. 2, 233-247.

\bibitem[Papin et al., 2002]{Papin2002}Papin, J.A., Price, N.D.,
Palsson, B.Ø., 2002. Extreme pathway lengths and reaction participation
in genome-scale metabolic networks. Genome Res. 12, 1889-1900.

\bibitem[Petzold et al., 2003]{Petzold2003}Petzold, A., Baker, D.,
Pryce, G., Keir, G., Thompson, E.J., Giovannoni, G., 2003. Quantification
of neurodegeneration by measurement of brain-specific proteins. J.
Neuroimmunology. 138(1), 45-48.doi:10.1016/j.jneumeth.2007.01.001

\bibitem[Preda et al., 2005]{Preda2005}Preda, A., Turetschek, K.,
Heike, D., Floyd, E., Viktor, N., Shames, D., Roberts, T., Carter,
W.O., Brasch, R., 2005. The Choice of Region of Interest Measures
in Contrast-Enhanced Magnetic Resonance Image Characterization of
Experimental Breast Tumors. Investigative Radiology. Breast Imaging
Part 1. 40(6), 349-354.

\bibitem[Rasmussen and Barrett, 1995]{Rasmussen1995}Rasmussen, S.,
Barrett, C., 1995. Elements of a Theory of Simulation. arXiv:adap-org/9504003v1.

\bibitem[Regan and Burgman, 2002]{Regan2002}Regan, H., Colyvan, M.,
Burgman, M., 2002. A taxonomy and treatment of uncertainty for ecology
and conservation biology. Ecol. Appl. 12(2), 618\textendash{}628.

\bibitem[Ricard, 2004]{Ricard2004}Ricard, J., 2004. Reduction, integration
and emergence in biochemical networks. Biol. of the Cell. 96, 719\textendash{}725.

\bibitem[Roman et al., 1998]{Roman1998}Roman-Roldan, R., Bernaola-Galvan,
P., Oliver, J., 1998. Sequence compositional complexity of DNA through
an entropic segmentation method. Phys. Rev. Lett. 80(6), 1344-1347.

\bibitem[Sasai et al., 1995]{Sasai1995}Sasai, Y, Lu, B, Steinbeisser,
H, De Robertis, E.M., 1995. Regulation of neural induction by the
chd and Bmp-4 antagonistic patterning signals in Xenopus. Nature.
376, 333\textendash{}336.

\bibitem[Seth, 2000]{Seth2000}Seth, A., 2000. On the relations between
behaviour, mechanism and environment : explorations in artificial
evolution. D. Phil. thesis, University of Sussex.

\bibitem[Siegfried, 2006]{Siegfried2006}Siegfried, K., 2006. Fishery
management in data-limited situations : applications to stock assessment,
marine reserve design and fish bycatch policy; PhD Thesis, University
of California, Santa Cruz.

\bibitem[Solomonoff, 1964]{Solomonoff1964}Solomonoff, R., 1964. A
Formal theory of Inductive Inference. Information and Control. 7,
1-22, 224-54.

\bibitem[Suzuki et al., 1995]{Suzuki1995}Suzuki, A., Shioda, N.,
Ueno, N., 1995. Bone morphogenic protein acts as a ventral mesoderm
modifier in early Xenopus embryos. Develop. Growth Differ. 37, 581\textendash{}588.

\bibitem[Szathmary et al., 2001]{Szathmary2001}Szathmáry, E., Jordán,
F., Pál, C., 2001. Can Genes Explain Biological Complexity? Science.
292(5520), 1315-1316.

\bibitem[Tautz et al., 1986]{Tautz1986}Tautz, D, Trick, M, Dover,
G.A., 1986. Cryptic simplicity in DNA is a major source of genetic
variation. Nature. 322, 652-656.

\bibitem[Terefework et al., 1998]{Terefework1998}Terefework, Z.,
Nick, G., Suomalainen, S., Paulin, L., Lindstrom, K., 1998. Phylogeny
of Rhizobium galegae with respect to other rhizobia and agrobacteria.
Int. J. Syst. Bacteriol. 48, 349\textendash{}356.

\bibitem[Testa and Kier, 2000]{Testa2000}Testa, B., Kier, L., 2000.
Emergence and Dissolvence in the Self-organisation of Complex Systems.
Entropy. 2, 1-25.

\bibitem[Thomas, 1971]{Thomas1971}Thomas, C., 1971. The genetic organization
of chromosomes. Ann. Rev. Genetics. 5, 237\textendash{}256.

\bibitem[Truman, 1973]{Truman1973}Truman, J., 1973. Physiobiology
of insect ecdysis III. Relationship between the hormonal control of
eclosion and of tanning in the tobacco hornworm manduca sexta; J.Exp.
Biol. 58, 821-829.

\bibitem[Tylianakis, 2008]{Tylianakis2008}Tylianakis, J.M., 2008.
Understanding the Web of Life: The Birds, the Bees, and Sex with Aliens.
PLoS Biol. 6(2): e47.

\bibitem[Uso et al., 2000]{Uso2000}Uso-Domenech, J., Villacampa-Esteve,
Y., Mateu-Mahiques, J., Sastre-Vazquez, P., 2000. Uncertainty and
complementarity principles in ecological models. Cybernetics and Systems.
31(2), 137-159.

\bibitem[Wahlsten et al., 2003]{Wahlsten2003}Wahlsten, D., Metten,
P., Phillips, T.J., Boehm, S.L. 2nd, Burkhart-Kasch, S., Dorow, J.,
Doerksen, S., Downing, C., Fogarty, J., Rodd-Henricks, K., Hen, R.,
McKinnon, C.S., Merrill, C.M., Nolte, C., Schalomon, M., Schlumbohm,
J.P., Sibert, J.R., Wenger, C.D., Dudek, B.C., Crabbe, J.C., 2003.
Different data from different labs: Lessons from studies of gene-environment
interaction. J. Neurobiol. 54, 283-311.

\bibitem[Wan and Wootton, 2000]{Wan2000}Wan, H., Wootton, J., 2000.
A global compositional complexity measure for biological sequences:
AT-rich and GC-rich genomes encode less complex proteins. Comput.
Chem. 24, 71-94.

\bibitem[Williams et al., 1981]{Williams1981}Williams, J.M., Peterson,
R.G., Shea, P.A., Schmedtje, J.F., Bauer, D.C., Felten, D.L., 1981.
Sympathetic innervation of murine thymus and spleen: evidence for
a functional link between the nervous and immune systems. Brain Res.
Bull. 6, 83-94.

\bibitem[Wilson and Hemmati, 1995]{Wilson1995}Wilson, P., Hemmati-Brivanlou,
A., 1995. Induction of epidermis and inhibition of neural fate by
Bmp-4. Nature. 376, 331\textendash{}333.

\bibitem[Wolpert and MacReady, 1997]{Wolpert1997}Wolpert, D., MacReady,
W., 1997. Self-dissimilarity: An empirical measure of complexity.
Working Paper 97-12-087, Sante Fe Inst.

\bibitem[Wootton and Federhen, 1996]{Wootton1996}Wootton, J., Federhen,
S., 1996. Analysis of compositionally biased regions in sequence databases.
Methods Enzymol. 266, 554-571.

\bibitem[Xu et al., 2006]{Xu2006}Xu, F., Xu, J., Tse, F., Tse, A.,
2006. Adenosine stimulates depolarization and rise in cytoplasmic
Ca2+ concentration in type I cells of rat carotid bodies. Am. J. Physiol.
Cell Physiol. 290, C1592-C1598.

\bibitem[Xu et al., 1995]{Xu1995}Xu, R.H., Kim, J., Taira, M., Zhan,
S., Sredni, D., Kung, H.F., 1995. A dominant negative bone morphogenetic
protein 4 receptor causes neuralization in Xenopus ectoderm. Biochem.
Biophys. Res. Comm. 212, 212\textendash{}219.

\bibitem[Young et al., 1991]{Young1991}Young, J.P.W., Downer, H.L.,
Eardly, B.D.,1991. Phylogeny of the phototropic rhizobium strain BTAi1
by polymerase chain reaction-based sequencing of a 16S rRNA gene segment.
J. Bacteriol. 173, 2271\textendash{}2277. .

\bibitem[Zhang et al., 2007]{Zhang2007}Zhang, Y., Zhang, Y., Chen,
J., Zhang, B., Pan, Y., Ren, L., Zhao, J., Luo, Y., Zhai, D., Wang,
S., Wang, J., 2007. Polybutylcyanoacrylate nanoparticles as novel
vectors in cancer gene therapy. Nanomedicine: Nanotech., Biol., and
Med. 3, 144\textendash{}153.20\textendash{}324.
\end{thebibliography}
\end{document}